\documentclass[twocolumn,tighten,times]{aastex63}

\usepackage{mathtools}
\usepackage{amsmath}
\usepackage{amssymb}
\usepackage{graphicx}
\usepackage{xcolor}
\usepackage{multirow}
\usepackage{soul}

\newcommand{\Msun}{\ifmmode {M_{\odot}}\else${M_{\odot}}$\fi} 
\newcommand{\Lsun}{\ifmmode {L_{\odot}}\else${L_{\odot}}$\fi} 
\newcommand{\Rsun}{\ifmmode {R_{\odot}}\else${R_{\odot}}$\fi} 
\newcommand{\Mb}{\ifmmode {M_{\text{bary}}}\else${M_{\text{bary}}}$\fi} 
\newcommand{\Mg}{\ifmmode {M_{\text{grav}}}\else${M_{\text{grav}}}$\fi} 

\shorttitle{EOS and neutrino effects in failed SNe}
\shortauthors{Ivanov \& Fern\'andez}

\begin{document}

\title{Mass ejection in failed supernovae: equation of state and neutrino loss dependence}

\author[0000-0001-7329-6963]{Mario Ivanov}
\affil{Department of Physics, University of Alberta, Edmonton, AB T6G 2E1, Canada.}

\author[0000-0003-4619-339X]{Rodrigo Fern\'andez}
\affil{Department of Physics, University of Alberta, Edmonton, AB T6G 2E1, Canada.}

\begin{abstract}
A failed core-collapse supernova from a non-rotating progenitor 
can eject mass due to a weakening of gravity associated to
neutrino emission by the protoneutron star. This mechanism yields
observable transients and sets an upper limit to the mass of the 
black hole (BH) remnant. Previous global simulations of this mechanism have included neutrino
losses parametrically, however, with direct implications for the
ejecta mass and energy. Here we evolve the 
inner supernova core with a spherically-symmetric, general-relativistic
neutrino radiation-hydrodynamic code until BH formation. We then
use the result in a Newtonian code that follows the response
of the outer layers of the star to the change in gravity and resolves
the surface pressure scale height. 
We find that the dense-matter equation of state (EOS) can introduce a factor $\sim 2$ variation 
in gravitational mass lost to neutrinos, with a stiff EOS matching previous parametric results, 
and a soft EOS yielding lower ejecta masses and energies by a factor of several. 
This difference is caused primarily by the longer time to BH formation in 
stiffer EOSs. With a soft EOS, our red and yellow supergiant progenitors 
fail to unbind mass if hydrogen recombination energy is not included.
Using a linear ramp in time for mass-energy lost to neutrinos
(with suitable parameters) yields a stellar response 
within $\sim 10\%$ of that obtained using the detailed history of neutrino losses. 
Our results imply quantitative but not qualitative modifications to previous predictions for
shock breakout, plateau emission, and final BH masses from these events.
\end{abstract}



\section{Introduction}
\label{s:intro}

Core-collapse supernova  theory has focused for decades on achieving successful
explosions in order to account for the majority of observed events (e.g.,
\citealt{janka_2016,burrows_2020}). Black hole (BH)
formation in failed supernovae has received more recent attention, due in
part to the need to explain the presupernova progenitor population (e.g.,
\citealt{kochanek_2008,smartt_2009}) and the mass distribution of remnant BHs
and neutron stars (NSs) (e.g., \citealt{raithel_2018,woosley_2020}). The increasing number of BH-BH
binaries detected in gravitational waves by Advanced LIGO \& Virgo
\citep{ligo_gwtc-1,ligo_gwtc-2} has further expanded our empirical knowledge of
the compact object mass distribution, with BH masses now extending to the
regime where pair-instability supernovae are relevant (with GW190521;
\citealt{ligo_gw190521}). On the theoretical side, simulations of BH-forming
supernovae -- that include some form of neutrino physics -- have been carried
out for large numbers of progenitors in spherical symmetry (e.g.,
\citealt{oconnor_2011,ugliano_2012,ertl_2016,ebinger_2019,warren_2019,couch_2020}),
and more recently in three-dimensions for a handful of cases (e.g.,
\citealt{chan_2018,kuroda_2018,walk_2019,chan_2020,pan_2020}). 

Observational predictions for BH formation events are strongly dependent on the
degree of rotation in the progenitor star. For stars with the right angular
momentum profile, a neutrino-cooled disk can form outside the BH and a
long-gamma-ray burst with an associated supernova can be produced (e.g.,
\citealt{macfadyen1999}, see also \citealt{nagataki_2018} for a review). Even
if the disk forms at larger radii, other transients could also arise due 
to outflows from the disk and/or thermonuclear explosions (e.g.,
\citealt{bodenheimer_1983,woosley_2012,kashiyama_2018,zenati_2020}).

In the absence of any significant rotation, mass ejection can still occur in a
failed supernova if a protoneutron star is formed (i.e., for progenitor masses
below the pair instability threshold). Neutrino emission reduces the
gravitational mass, generating an imbalance between the acceleration of gravity
and the pressure gradient in the progenitor, driving an outflow
\citep{Nadyozhin_1980}. Mass ejection from this mechanism {therefore} sets an
upper limit on the mass of the BH, since any subsequently formed accretion disk
would eject additional mass, further reducing that of the remnant BH.  For red
supergiants (RSGs), this mechanism can eject the weakly bound hydrogen
envelope, generating an optical transient with peak luminosity $\sim
10^{39}$\,erg\,s$^{-1}$ and $\sim $\,yr duration
\citep{Lovegrove_Woosley_2013}. The failed supernova candidate N6946-BH1,
associated with the disappearance of a RSG, showed a similar year-long
transient \citep{gerke_2015,adams_2017,basinger_2020}.

This neutrino-induced mass ejection is not limited to RSGs, however.
\citet[hereafter F18]{F18} showed that blue supergiants (BSGs) and Wolf-Rayet
(WR) stars also eject mass through this mechanism, although the ejecta masses,
energies, and velocities differ significantly from those in RSGs. This diversity
of progenitors implies electromagnetic signals that span a wide range in
luminosity, duration, and peak wavelength, resulting in new types of transients.
Of particular recent interest are failed supernovae from BSGs as possible
progenitors of fast blue optical transients (e.g., \citealt{kashiyama_2015,margutti_2019}) and 
very massive BH remnants from failed supernovae around the lower edge 
of the ``pair instability gap"
(e.g., \citealt{farmer_2019,mapelli_2020,marchant_2020,renzo_2020}).

A key limitation of the global simulations of F18 is the use of an analytic model to
describe the decrease in gravitational mass with neutrino emission.  While this
parametric approach, which follows that of \citet{Lovegrove_Woosley_2013},
provides a computationally inexpensive input, it takes
important quantities (neutrino cooling timescale, binding energy and maximum
mass of the {NS}) as free parameters.  The physics that sets these
quantities is complex, depending on the transport of neutrinos in the
protoneutron star and on the equation of state (EOS) of dense matter.

Here we improve upon the work of F18 by modeling the evolution of the inner
core with the spherically symmetric, general-relativistic neutrino
radiation-hydrodynamics code {\tt GR1D} \citep{oconnor_2010}, which yields the
history of gravitational mass lost to neutrinos until BH formation given a
dense-matter EOS and presupernova star.  This physically-grounded
gravitational mass decrease is then used, in lieu of a parametric scheme, as an
input to the same global hydrodynamic setup used by F18 to characterize the
response of the star to the change in gravity. {These global} simulations
are performed at spatial resolutions such that the
surface pressure scale height is resolved in all progenitors, thereby also
improving upon the work of F18.  We focus on understanding the dependence of
the ejecta {properties} on the EOS, on the spatial resolution of the
simulation, on the need to know the detailed neutrino emission history of the
protoneutron star, and on the implications of these inputs for the
electromagnetic signatures expected from these events.

The structure of this paper is the following. Section~\ref{s:methods} describes
our assumptions and the numerical method employed, the presupernova progenitors
used, and the range of models evolved. Section~\ref{s:results} provides an
overview of results for various progenitors, the effect of varying the EOS, the
sensitivity to the history of gravitational mass loss, the effects of spatial
resolution, and the observational implications. A summary and discussion
follows in Section~\ref{s:summary}.

\section{Methods}
\label{s:methods}

\subsection{Physical Model and Approximations}

\begin{figure}
\includegraphics[width=\columnwidth]{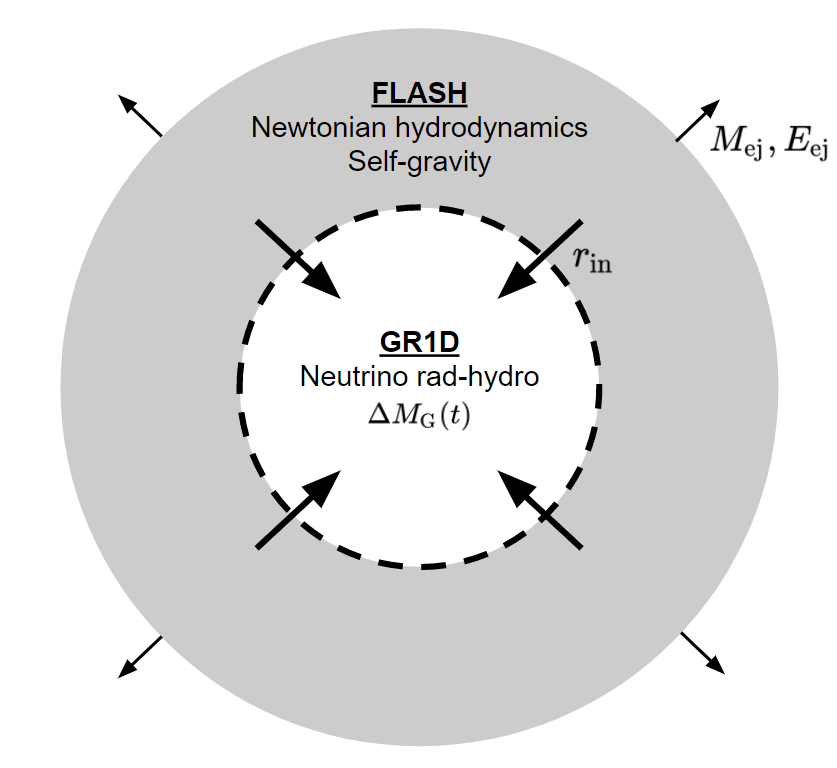}
\caption{
Schematic of our computational approach. For a 
given presupernova progenitor (\S\ref{s:progenitors}) and dense-matter EOS, the evolution of the 
inner core is followed with the general-relativistic,
neutrino radiation-hydrodynamics code {\tt GR1D} (\S\ref{s:inner_core}), 
focusing on the gravitational mass lost to neutrinos $\Delta M_{\rm G}(t)$ [eq.~\ref{eq:dM_definition}]
at some transition radius $r_{\rm in}$. The region outside $r_{\rm in}$
is then evolved with {\tt FLASH}, accounting for the mass flowing supersonically
into $r_{\rm in}$ as well as the change in gravity due to $\Delta M_{\rm G}(t)$ (\S\ref{s:flash_methods}),
and focusing on any mass ejected from the stellar surface. The majority
of our models only interpolate $\Delta M_{\rm G}(t)$ from {\tt GR1D}, while
a smaller sample employs an analytic approximation for this function or
maps the initial condition for {\tt FLASH} directly from {\tt GR1D} (Table~\ref{t:evolution_modes}).
}
\label{f:schematic}
\end{figure}

Our aim is to compute the hydrodynamic response of the star to the decrease in
the gravitational mass from neutrino emission after core bounce in the case where
the supernova fails and a {BH} forms. We restrict ourselves to 
progenitors for which rotation is negligible, and ignore the effect of hydrodynamic
instabilities that operate within the first second after bounce, which have
an important role in a successful explosion (e.g., \citealt{janka_2016}; here we only consider failures).
We therefore carry out our analysis in spherical symmetry. We also ignore any 
circumstellar material that the star could have ejected prior to undergoing core-collapse.
This material can modify the observational signatures of shock breakout 
(e.g., \citealt{chevalier_2011,katz_2012,haynie_2020}). 
Our focus is the total energy and mass of any ejecta arising from the change 
in gravitational acceleration.

In previous work (F18), we removed the inner stellar core and 
computed the hydrodynamic response of the star
using analytic prescriptions for the evolution of the gravitational mass
enclosed within this radius. This enclosed mass evolves due to material accreted through
the inner radius and due to neutrino emission during the protoneutron star phase (see also
\citealt{Lovegrove_Woosley_2013}). This modeling approach is acceptable because material that accretes
through the inner boundary reaches supersonic speeds shortly after collapse, 
without any pressure feedback on the outer layers.

Here we replace the analytic prescriptions for the inner core evolution
used in F18 with a neutrino radiation-hydrodynamic calculation (Figure~\ref{f:schematic}). Due to
the complexity of the latter and the limitations in thermodynamic range 
set by publicly available EOSs of nuclear matter, we can carry out this calculation 
self-consistently only until a BH forms, and within a limited volume inside the star. 
Nevertheless, the supersonic character of matter infall and the rapidly increasing 
dynamical time with stellar radius allows for a decoupling in the modeling of the
inner core and the outer layers of the star with separate codes for
each regime, without a significant loss in consistency. 

\subsection{Pre-supernova progenitors}
\label{s:progenitors}

The pre-supernova stars we explore are shown in Table~\ref{t:progenitors}. 
We adopt the same fiducial solar-metallicity progenitors as in F18: a 
15\,$M_\odot$ RSG (denoted by R15), a 25\,$M_\odot$ BSG (B25), and a 40\,$M_\odot$ WR (W40). 
These models cover three different regimes in core compactness parameter 
\citep{oconnor_2011,sukhbold_2014,sukhbold_2018}
\begin{equation}
\label{eq:compactness}
\xi_{\rm 2.5} = \frac{2.5}{r(M=2.5M_\odot)/1000 \ \rm km},
\end{equation}
and in envelope compactness
\begin{equation}
\label{eq:global_compactness}
\xi_{\rm env} = \frac{(M_{\rm cc}/M_\odot)}{(R_{\rm cc}/R_\odot)},
\end{equation}
where $M_{\rm cc}$ and $R_{\rm cc}$ are the mass and radius of the star at
core-collapse. The values of $\xi_{\rm 2.5}$ and $\xi_{\rm env}$ are 
shown in Table~\ref{t:progenitors} (see \citealt{pejcha_2015,ertl_2016,mueller_2016,murphy_2017} for
explodability criteria other than $\xi_{\rm 2.5}$).

In addition to these baseline models, we consider two yellow supergiants (YSGs), one of 
22\,$M_\odot$ and solar metallicity (Y22), and another of 25\,$M_\odot$
and low-metallicity (Y25);
a massive 80\,$M_\odot$ low-metallicity blue supergiant (B80); and a 50\,$M_\odot$ 
solar-metallicity WR (W50). The YSGs are such than one of 
$\{\xi_{\rm 2.5},\xi_{\rm env}\}$ has a value similar to R15, while
the other varies. W50 has a higher core-compactness than W40, 
and B80 has high values of both core- and envelope compactness 
(in F18 no mass is ejected from this model).
All of these progenitors are computed with the stellar evolution code \texttt{MESA}
version 6794 \citep{paxton2011,paxton2013,paxton2015,paxton2018}.
Parameters and physical choices are described in \citet{F18} 
(see also \citealt{fuller_2015}), and inlists are publicly 
available\footnote{\url{bitbucket.org/rafernan/bhsn\_mesa\_progenitors}}.

To connect with previous work, we also evolve models s20 and s40 from \citet{woosley_2007}.
These progenitors have been used in {BH} 
formation simulations (e.g., \citealt{oconnor_2015,pan_2018})
and in a code-comparison study of 1D core-collapse supernova simulations \citep{oconnor_2018},
providing calibration values.

\begin{table}
\centering
\caption{Presupernova progenitors used in this study. Columns from left to right show model
name, type of star 
(RSG: red supergiant, YSG: yellow supergiant, BSG: blue supergiant, WR: Wolf-Rayet),
zero-age main sequence mass, 
initial metallicity, mass at core-collapse, effective temperature at core-collapse, core compactness
(eq.~[\ref{eq:compactness}]), and envelope compactness (eq.~[\ref{eq:global_compactness}]).
Model S40 has the structure of a BSG but it is truncated below its photosphere.
}
\label{t:progenitors}
\begin{tabular}{lccccccc}
\hline
\noalign{\smallskip}
Model & Type & M$_{\text{zams}}$ & $Z$ & M$_{\text{cc}}$ & 
T$_{\text{eff}}$ & $\xi_{2.5}$ & $\xi_{\text{env}}$ \\
 & & (\Msun) & ($Z_\odot$) & (\Msun) & ($10^3$ K) \\ 
\noalign{\smallskip}
\hline
\noalign{\smallskip}
R15 & RSG & 15 & 1.00 & 11   & 3   & 0.24  & 0.010 \\
B25 & BSG & 25 & 1.00 & 12   & 15  & 0.33  & 0.120 \\
W40 & WR  & 40 & 1.00 & 10   & 260 & 0.37  &  27   \\
\noalign{\smallskip}
Y22 & YSG & 22 & 1.00 &  11  & 5   & 0.54 & 0.016 \\
Y25 & YSG & 25 & 0.01 &  23  & 4.6 & 0.25 & 0.024 \\
W50 & WR  & 50 & 1.00 &  9   & 215 & 0.55 &   22  \\
B80 & BSG & 80 & 0.01 &  55  &  28 & 0.97 & 0.79  \\
\noalign{\smallskip}
S20 & RSG & 20 & 1.00 & 16   & 2.5 & 0.28 & 0.015 \\
S40 & BSG & 40 & 1.00 & 15   & ... & 0.54 & 1.3   
\end{tabular}
\end{table}

\subsection{Inner core evolution}
\label{s:inner_core}

The evolution of the inner stellar core from collapse until BH
formation is modeled with the spherically-symmetric
neutrino radiation-hydrodynamics code {\tt GR1D}\footnote{Available at \url{stellarcollapse.org}\label{note_collapse}} 
version 1 \citep{oconnor_2010}.
The code solves the equations of general-relativistic hydrodynamics in spherical 
coordinates using the radial gauge, polar slicing metric \citep{romero_1996}.
The finite-volume hydrodynamics solver employs a piecewise-parabolic reconstruction,
and performs a temporal update using a second-order Runge-Kutta scheme. 

During collapse, neutrino effects are modeled via a parameterization of the electron 
fraction with density \citep{liebendoerfer_2005}. After bounce,
neutrino cooling in this version of {\tt GR1D} is modeled with a gray leakage scheme 
for $\nu_{\rm e}$, $\bar\nu_{\rm e}$ and a composite heavy lepton neutrino $\nu_{\rm x}$. 
The opacity has contributions from charged-current absorption on nucleons and neutral-current 
scattering on nucleons and nuclei. Emission accounts for charged-current reactions {as well as} thermal 
emission from {electron-positron} pair annihilation and plasmon decay 
{(we did not include nucleon-nucleon bremsstrahlung)}. The local effective emission rate 
is an interpolation between the diffusive and free emission rates \citep{ruffert_1996,rosswog_2003}. 
Neutrino heating due to charge current absorption is computed from the enclosed local luminosity obtained
from the leakage scheme. The local outgoing luminosity is then corrected for the neutrino energy
lost to heating.

We consider three finite-temperature EOSs 
in our calculations: SFHo \citep{steiner_2013} as our default (soft) case,
DD2 \citep{hempel_2012} as a stiff variant, and LS220 \citep{lattimer_1991} as another
reference case. These EOSs are commonly used in the core-collapse supernova and NS merger literature 
(e.g., \citealt{pan_2018,vincent_2019}),
thus providing a connection to previous work (while DD2 and SFHo are consistent
with experimental constraints and unitary gas bounds on the symmetry energy and its
density derivative, LS220 is not; c.f. \citealt{tews_2017}).
We neglect light clusters when computing opacities and emissivities with the DD2 and SFHo EOSs, 
as these species have a sub-dominant effect on the post-bounce evolution
\citep{yudin_2019,nagakura_2019}.

In most models, the computational domain is discretized with a uniform grid of $200$ cells from the origin out 
to $20$\,km, and logarithmic spacing for larger radii, for a total grid size of $1000$ cells. 
In all RSG models and some YSGs and BSGs, we double the resolution in the uniform section of the grid ($r<20$\,km) as
these models take longer to reach BH formation and reach more compact shock radii.
The maximum radius is set by a density close to the lowest value in the tabulated EOS 
($2\times 10^3$\,g\,cm$^{-3}$), corresponding typically to several times $10^9$\,cm, much smaller
than the radius of the star at core-collapse. The dynamical time at the outer boundary is typically
$\sim 10$\,s, much longer than the time to form a BH in most models, 
justifying our approximation of neglecting the evolution of the outer stellar layers 
when evolving the core with {\tt GR1D}. 
Simulations are deemed to have formed a BH when the density increases rapidly with time
to values $\sim 1-3\times 10^{15}$\,g\,cm$^{-3}$ (near the upper limit of the EOS table), 
at which point the simulation stops.
In only one case (model R15 with the DD2 EOS) we fail to reach BH formation within $4.37$\,s
of evolution and the code crashes, although based on the central value of the lapse function ($0.44$), the
model is close to BH formation. 

While a newer version of {\tt GR1D} is available, which treats neutrinos with a multigroup
moment (M1) scheme \citep{oconnor_2015} and therefore provides a more accurate measure of the
gravitational mass 
lost to neutrinos, the convergence of the transport algorithm near {BH} 
formation is more
fragile than for the leakage scheme. Evolving the s40 progenitor with the LS220 EOS, we obtain
post-bounce times to BH formation $\{0.561,0.556\}$\,s, maximum PNS baryonic masses
(at a density of $10^{12}$\,g\,cm$^{-3}$) $\{2.448,2.372\}\,M_\sun$, 
and maximum PNS gravitational masses $\{2.309,2.246\}M_\sun$
with the leakage and M1 versions of {\tt GR1D}, respectively. These masses and BH formation
times are the same as those reported in \citet{oconnor_2015} for the leakage version, 
while for the M1 version the masses are the same within $1\%$ but the BH
formation time is about $4\%$ longer. These results are also consistent 
(within $\sim 10\%$) with the 1D results of \citet{pan_2018} 
for the s40 progenitor using the LS220 EOS and a different code.

\subsection{Response of the star to neutrino losses}
\label{s:flash_methods}

The hydrodynamic response of the outer layers of the star to the decrease in
gravity due to neutrino emission
is modeled in spherical symmetry with the Newtonian hydrodynamic
code {\tt FLASH} version 3 \citep{fryxell00, dubey2009}, modified as 
described in \citet{F12} and F18. 
The code solves the Euler equations in 
spherical coordinates with the dimensionally-split Piecewise Parabolic Method 
(PPM; \citealt{colella84}, \citealt{fryxell1989}) 
and the \textit{Helmholtz} EOS \citep{timmes2000}. 

The computational domain spans a radial interval $[r_{\rm in},r_{\rm out}]$ that varies
for different evolution modes and progenitors, as explained below. The boundary conditions 
are set to outflow at $r=r_{\rm in}$ and $r=r_{\rm out}$. 
The mass flowing out through the inner radial boundary is kept track of 
as a scalar baryonic mass $M_{\rm B,flash}$,
such that total mass is conserved close to machine precision (F18).

\begin{figure}
\includegraphics*[width=\columnwidth]{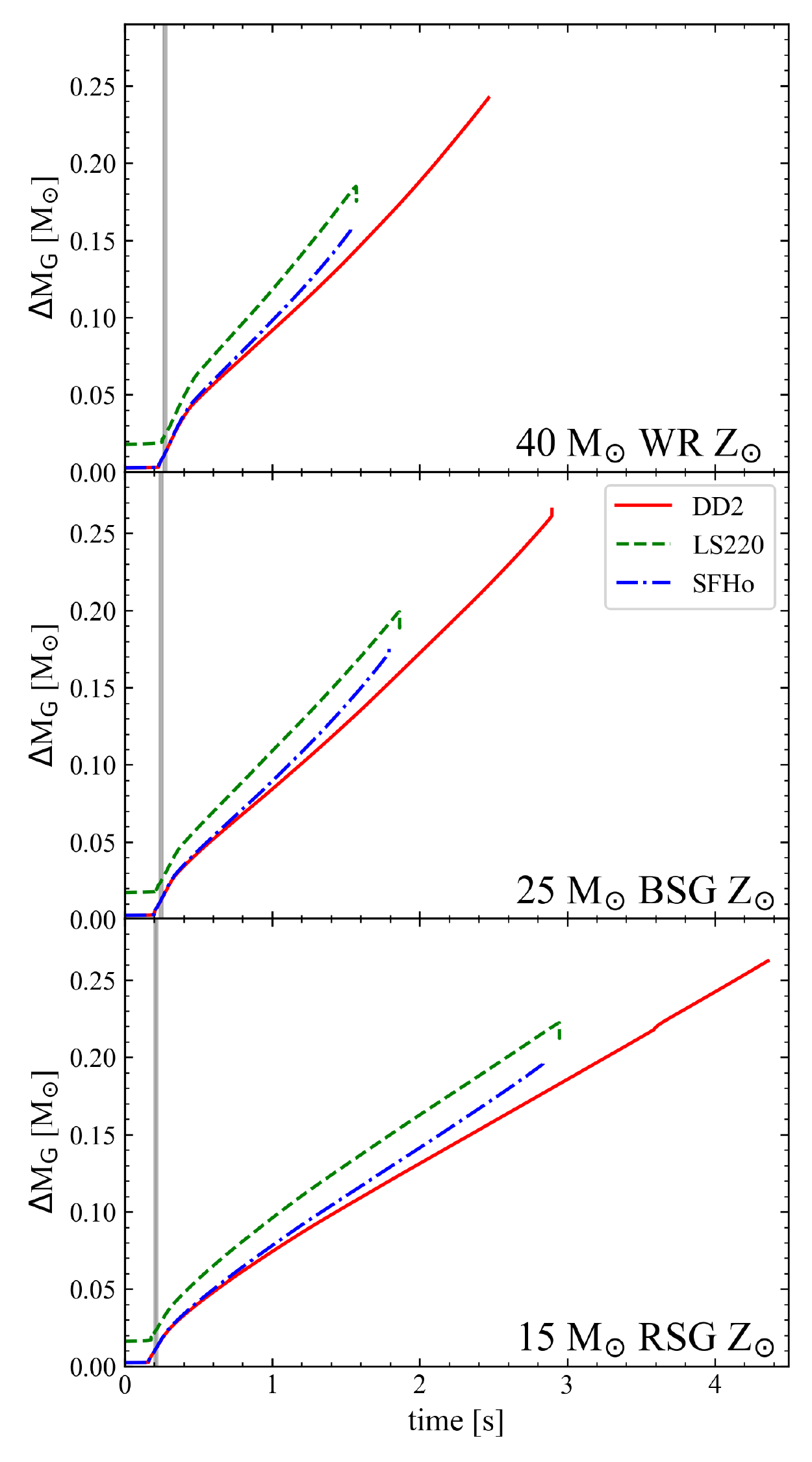}
\caption{
Evolution of the difference $\Delta M_{\rm G}$ (eq.~[\ref{eq:dM_definition}])
between baryonic and gravitational masses in {\tt GR1D}, evaluated at
{$r=2\,000$}\,km, as a function of time from the onset of core-collapse. Top,
middle, and bottom panels correspond to our fiducial presupernova models W40,
B25, and R15, respectively (Table~\ref{t:progenitors}).  For each case, the
result obtained with the DD2, SFHo, and LS220 EOS is shown, as labeled.  The
gray vertical band shows the range of maximum shock radius time $t_{\rm stall}$
obtained with the three EOSs ($0.204-0.220$\,s for R15, $0.236-0.257$\,s for
B25, and $0.261-0.286$\,s for W40, with the largest values corresponding to the DD2 EOS).
{The larger value of $\Delta M_{\rm G}$ at $t=0$ (eq.~\ref{eq:dM_offset_initial}) for the LS220 EOS 
is due to an offset in the internal energy needed to connect the high- and 
low-density regimes of the EOS \citep{oconnor_2010}.}
}
\label{f:dM_t}
\end{figure}

The gravitational acceleration in {\tt FLASH} $\mathbf{g}_{\rm F}$ is a sum
of the contribution from the baryonic mass in the computational domain
and the gravitational mass $M_{\rm G,flash}$ inside $r_{\rm in}$,
\begin{equation}
\mathbf{g}_{\rm F}(r,t) = -\frac{G}{r^2}\left[M_{\rm G,flash} + 4\pi\int_{r_{\rm in}}^r 
\rho(\zeta,t)\, \zeta^2 d\zeta\right]\hat r,
\end{equation}
{where $\rho$ is the mass density.}
Initially, the gravitational mass $M_{\rm G,flash}$ is either equal to the baryonic mass enclosed 
within $r=r_{\rm in}$ in the presupernova progenitor, or otherwise mapped from {\tt GR1D},
depending on the mode of evolution as described below. This gravitational mass is subsequently updated from 
time $t_n$ to $t_{n+1}$ by adding the change due to the baryonic mass
flowing through the inner boundary over the time step, and correcting for the instantaneous 
difference between baryonic and gravitational masses as computed by {\tt GR1D} or from 
an analytic fit,
\begin{eqnarray}
\label{eq:mg_flash_evolution}
M_{\rm G,flash}^{(n+1)} & = & M^{(n)}_{\rm B,flash}  
                    + 4\pi\int_{t_n}^{t_{n+1}} \left[r^2 \rho \max(-v_r,0)\right]\big|_{r_{\rm in}} dt\nonumber\\
                       && - \Delta M_{\rm G}(t_{n+1}) + \Delta M_{\rm G}(t=0),
\end{eqnarray}
where
\begin{equation}
\label{eq:dM_definition}
\Delta M_{\rm G}(t_n) = M_{\rm B,gr1d}(t_n) - M_{\rm G,gr1d}(t_n),
\end{equation}
is the instantaneous difference between the baryonic and gravitational masses
enclosed by $r=r_{\rm in}$ in {\tt GR1D} ($M_{\rm B,gr1d}$ and $M_{\rm G,gr1d}$, respectively),
and $\Delta M_{\rm G}(t=0)$ is an initial {EOS-dependent} offset 
between these two masses in {\tt GR1D} 
\begin{equation}
\label{eq:dM_offset_initial}
\Delta M_{\rm G}(t=0) \simeq 4\pi\int_0^{r_{\rm in}} \rho r^2\,dr
       \left[\frac{GM_{\rm G,gr1d}(r)}{c^2 r} - \frac{e_{\rm int}}{c^2} \right],
\end{equation} 
where $e_{\rm int}$ is the specific internal energy, and we have assumed
non-relativistic collapse speeds as well as $GM_{\rm G,gr1d}(r)/c^2 \ll r$ in
deriving equation~(\ref{eq:dM_offset_initial}). For the
SFHo and DD2 EOSs, the term in square brackets is $\sim 10^{-3}$, whereas for
LS220 it is $\sim 0.01$, given an extra offset in $e_{\rm int}$ (of order the
nuclear binding energy per nucleon) needed to connect the nuclear EOS with a
low-density continuation \citep{oconnor_2010}. Figure~\ref{f:dM_t} shows the
evolution of $\Delta M_{\rm G}$ at $r_{\rm in} = 2\,000$\,km in {\tt GR1D} for our fiducial
progenitors using all three EOSs considered in this study.

Since at the onset of significant neutrino emission we have 
$M_{\rm G,flash}\simeq M_{\rm G,gr1d}\simeq M_{\rm B,gr1d}$, 
{the} formulation {in equation~(\ref{eq:mg_flash_evolution})} 
preserves consistency in the mass evolution within {\tt FLASH}, which is
entirely baryonic, while also accounting for the mass-energy lost to neutrinos via
$\Delta M_{\rm G}$. 
When a BH forms at time $t=t_{\rm bh}$, the mass difference 
$\Delta M_{\rm G}(t_{\rm bh})$ becomes a constant in equation~(\ref{eq:mg_flash_evolution}).

The initial condition for {\tt FLASH} and the evolution of the inner core ($r<r_{\rm in}$) are 
treated in three different ways, to assess the sensitivity of our results to the details of the
inner core history (Table~\ref{t:evolution_modes}):

\begin{table}
\caption{Evolution modes for {\tt FLASH} simulations.}
\centering
\label{t:evolution_modes}
\begin{tabular}{lccc}
\hline
     & \multicolumn{2}{c}{$\Delta M_{\rm G}(t)$} & initial condition\\
     &  from {\tt GR1D}  & analytic & from {\tt GR1D}        \\
\hline
Interpolation (1) & yes  & no & no \\
Analytic      (2) & no  & yes & no \\
Remap         (3) & yes  & no & yes \\
\end{tabular}
\end{table}

\begin{figure}
\includegraphics*[width=\columnwidth]{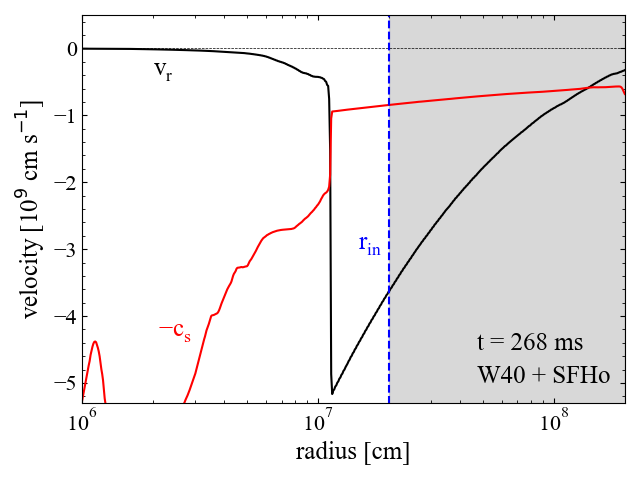}
\caption{
Velocity and sound speed as a function of radius in the core-collapse
simulation of the W40 progenitor carried out with {\tt GR1D} using the SFHo EOS.  
The time shown corresponds to that at which, when using \emph{remap} 
evolution mode (Table~\ref{t:evolution_modes}), we map variables into {\tt FLASH} for
subsequent evolution. The vertical blue line indicates the position of the inner radial 
boundary in {\tt FLASH}. 
}
\label{f:velx_csnd_rcut}
\end{figure}

\begin{enumerate}
\item \emph{Interpolation}: by default, we initialize {\tt FLASH} with the pre-collapse profile, 
and interpolate $\Delta M_{\rm G}(t)$ at $r=r_{\rm in}$ as a function of time from {\tt GR1D}.
This approach provides a more realistic value for the gravitational mass
lost to neutrinos relative to the models of F18, while starting from the same initial condition.
Figure~\ref{f:dM_t} shows the evolution the mass difference $\Delta M_{\rm G}(t)$ at 
{$r=2\times 10^8$\,cm}
for the three fiducial progenitors and EOSs used. The curves start from a very small value 
$\Delta M_{\rm G}(t=0)$ and increase almost linearly until BH formation. The inner radial boundary 
for these models is located at $r_{\rm in}=2\times 10^8$\,cm as in F18 (approximately at the outer edge 
of the iron core), and $\Delta M_{\rm G}(t)$ at that radius is interpolated for use in {\tt FLASH}.

\item \emph{Analytic}: given the overall simplicity of the function $\Delta M_{\rm G}(t)$, we evolve a second
group of models by initializing the domain in the same way as with Interpolation mode, but now
we parameterize the function $\Delta M_{\rm G}(t)$ as a linear ramp
that turns on and off at specified times $t_{\rm stall}$ and $t_{\rm bh}$, respectively, reaching the same maximum
value as the instantaneous function $\Delta M_{\rm G}(t)$. 
The aim is to quantify the degree to which the details of the gravitational mass loss history
(as opposed to just the final magnitude and overall timescale) influences the results.
The inner boundary for these models is also located at $r_{\rm in}=2\times 10^8$\,cm.

\item \emph{Remap}: a third group of models are such that the initial condition for {\tt FLASH} is mapped from {\tt GR1D},
in addition to interpolating the mass difference $\Delta M_{\rm G}(t)$ at $r=r_{\rm in}$ as a function of time. 
Profiles of density, pressure, velocity, and composition are mapped at a time $t_{\rm stall}$ when the shock
reaches its maximum amplitude, usually $\sim$\,200 ms after the onset of collapse (Figure~\ref{f:dM_t}). 
The inner radius $r_{\rm in}$ of the computational domain in {\tt FLASH} is chosen such 
that (1) the shock radius in {\tt GR1D} never exceeds it, and (2) the flow at this radius is supersonic, 
so that there is no hydrodynamic feedback to regions outside this transition. Figure~\ref{f:velx_csnd_rcut}
shows a snapshot of the velocity profile in a {\tt GR1D} run of model W40 at the time of mapping into {\tt FLASH}.
For this model, $r_{\rm in} = 2\times 10^7$\,cm and the time of mapping is $\sim 270$\,ms after
the onset of collapse.
This is not our default mode of evolution because inconsistencies between general-relativistic
and Newtonian evolution are not entirely negligible at the level of precision needed to
model this mass ejection mechanism.
\end{enumerate}

The computational domain is discretized with a logarithmic grid using a resolution of $2\,048$ cells per decade 
in radius ($\Delta r / r \simeq 0.11\%$) for RSG, YSG, and BSG, models, and twice that value 
($\Delta r/r \simeq 0.06\%$) for WRs. At this resolution, the pressure scale heights at the surfaces 
of all progenitors are resolved. 
We also evolve models with a lower resolution of $512$ cells per decade in radius to compare with the
results of F18, which used this value in their highest resolution models.

Like in F18, following BH formation, the inner radial boundary of the simulation
is moved out by a factor of $10$ at specific times, once the
flow in the entire region $[r_{\rm in},10r_{\rm in}]$ has become supersonic, 
so that the minimum hydrodynamic time step becomes larger ($\Delta r \propto r$ in a logarithmic grid) 
and evolution of the shock up to the stellar surface and beyond can be followed at a smaller
computational cost.

The outer radius of the computational domain in {\tt FLASH} is set at
$r_{\rm out} = \{2\times 10^{16},2\times 10^{15},2\times 10^{14}\}${\,cm} for RSG/YSG, BSG, and WR
progenitors, respectively, corresponding to factors $30-1000$ times the stellar surface
at core collapse. 
The domain outside the star is filled with a constant-density ambient medium in hydrostatic
equilibrium, with the same composition as the stellar surface. The ambient densities are 
$\{10^{-18},10^{-16},5\times 10^{-13}\}${\,g\,cm$^{-3}$} for RSG/YSG, BSG, and WR progenitors, respectively. These densities
are low enough that the ambient mass swept up by the shock is much smaller ($\ll 1\%$) than
the ejecta mass itself, with negligible slowdown. While the mass in ambient for the RSG/YSG
models could in principle reach $\sim 0.1M_\odot$ at the maximum simulation radii, we normally
stop our simulations much earlier than that point. 
A floor of temperature at $10^4$\,K is adopted in all simulations, consistent with the low-temperature 
limit of the Helmholtz EOS in {\tt FLASH} (in practice this is what sets the stopping time for
RSG, YSG, and some BSG simulations). The density floor is set $100$ times lower than the ambient 
for RSGs, YSGs, and BSGs, and a factor of 5 lower than the ambient for WRs.

\subsection{Models Evolved}

All of our simulations are listed in Table~\ref{t:results}. We adopt
the SFHo EOS and interpolation of $\Delta M_{\rm G}(t)$ from {\tt GR1D} 
(Table~\ref{t:evolution_modes}) as our default setting. 

The three fiducial progenitors described in \S\ref{s:progenitors} 
(R15, B25, W40) are evolved using the three EOSs described 
in \S\ref{s:inner_core}, with progenitor names appended the letters \{S,L,D\} when using 
the SFHo, LS220, or DD2 EOS, respectively, and ending in \{1,2,3\} in accordance
to the inner core evolution modes listed in Table~\ref{t:evolution_modes}. 
For example, model R15S1 is the R15 progenitor evolved with the SFHo EOS in {\tt GR1D},
and interpolating $\Delta M_{\rm G}(t)$ from {\tt GR1D} into {\tt FLASH}.
We also evolve the 3 fiducial progenitors varying the evolution mode
of the inner core, using the SFHo EOS. The remaining progenitors are all 
evolved using the SFHo EOS and interpolation of $\Delta M_{\rm G}(t)$, with the 
exception of the B80 progenitor, which we evolve using the SFHo and DD2 EOS. 

Each progenitor is evolved at the maximum resolution listed in \S\ref{s:flash_methods},
while fiducial progenitors are also evolved at the lower resolution used in F18, 
to compare results. The maximum
run time for each model is set either by the shock emerging from the 
star and reaching nearly constant total energy, or otherwise when the temperature 
in the postshock flow reaches the floor value of $10^4$\,K, at which 
point non-conservation of thermal energy ensues and results become unreliable.

\section{Results}
\label{s:results}

\begin{table*}
\caption{Summary of results.
Columns from left to right show model name, evolution mode (I: interpolation,
A: analytic, R: remap, c.f. Table~\ref{t:evolution_modes}), EOS used in {\tt GR1D}, time to BH formation
from the onset of core-collapse $t_{\rm bh}$, maximum gravitational mass lost to neutrinos $\Delta M_{\rm G}(t_{\rm bh})$,
maximum kinetic energy of the acoustic pulse in the {\tt FLASH} simulation $E^{\rm sim}_{\rm k,max}$, 
mass ejected in the {\tt FLASH} run $M_{\rm ej}$, total energy of ejecta in {\tt FLASH} run $E_{\rm ej}$, 
and recombination energy $E_{\rm rec}$ (eq.~\ref{eq:Erec_definition}) for stars with
extended hydrogen envelopes. Results are reported for {\tt FLASH} runs at their maximum resolution, 
see \S\ref{s:resolution} for results at lower resolution.
Model W40S1.7 has its inner radial boundary at $r_{\rm in}=2\times 10^7$\,cm, 
instead of the default $r_{\rm in}=2\times 10^8$\,cm for \emph{interpolation} mode. 
}
\centering
\label{t:results}
\begin{tabular}{lcccccccc}
\hline
\noalign{\smallskip}
Model & Mode & EOS & $t_{\rm bh}$ (s) & $\Delta M_{\text{G}}$ ($M_\odot$) & $E^{\rm sim}_{\rm {k,max} }$ ($10^{47}$\,erg) & 
                         {$M_{\text{ej}}$ (\Msun)}  & $E_{\text{ej}}$ ($10^{47}$\,erg) & $E_{\rm rec}$ ($10^{47}$\,erg)\\
\noalign{\smallskip}
\hline
R15S1 & I & SFHo  & 2.836    & 0.196      & 2.15 &  2.19 & -0.119 & 0.398 \\ 
R15S2 & A &       &          &            & 2.06 &  2.16 & -0.154 & 0.388 \\ 
R15S3 & R &       &          &            & 0.57 &  0.99 & -0.268 & 0.179 \\ 
R15L1 & I & LS220 & 2.947    & 0.222      & 2.42 &  2.42 & -0.103 & 0.440 \\ 
R15D1 &   & DD2   & $>$4.359 & $>$0.262   & 3.59 &  3.37 &  0.489 & 0.612 \\ 
\noalign{\smallskip}
S20S1 &   & SFHo  & 2.188    & 0.173      & 0.69 &  0.70 & -0.320 & 0.128 \\ 
\noalign{\smallskip}
\hline
\noalign{\smallskip}
      &   &       &          & & & $M_{\text{ej}}$ ($10^{-2}$\,\Msun) & & \\
\noalign{\smallskip}
\hline
\noalign{\smallskip}
B25S1 & I & SFHo  & 1.791    & 0.173    & 2.27 & 2.80 & 0.399 & \\ 
B25S2 & A &       &          &          & 2.27 & 2.82 & 0.404 & \\ 
B25S3 & R &       &          &          & 0.14 & 0.45 & 0.012 & \\ 
B25L1 & I & LS220 & 1.864    & 0.198    & 2.55 & 3.18 & 0.593 & \\ 
B25D1 &   & DD2   & 2.895    & 0.261    & 5.40 & 5.45 & 1.76  & \\ 
\noalign{\smallskip}
Y22S1 &   & SFHo  & 0.931    & 0.139    & 0.55 & 5.39 & -0.013 & 0.010 \\ 
Y25S1 &   &       & 2.542    & 0.175    & 3.18 & 15.5 & -0.082 & 0.028 \\
\noalign{\smallskip}
\hline
\noalign{\smallskip}
        &   &        &       & & & $M_{\text{ej}}$ ($10^{-4}$\,\Msun) & & \\
\noalign{\smallskip}
\hline
\noalign{\smallskip}
W40S1   & I & SFHo   & 1.535 & 0.157  & 1.40 & 1.44 & 0.067 & \\ 
W40S2   & A &        &       &        & 1.36 & 1.36 & 0.063 & \\
W40S3   & R &        &       &        & 0.06 & 0.01 & $<$0.001 & \\ 
W40L1   & I & LS220  & 1.570 & 0.184  & 1.52 & 1.63 & 0.077 & \\
W40D1   &   & DD2    & 2.466 & 0.242  & 3.92 & 6.30 & 0.326 & \\ 
\noalign{\smallskip}
W40S1.7 &   & SFHo   & 1.535 & 0.157  & 1.68 & 1.91 & 0.090 & \\ 
W50S1   &   &        & 0.895 & 0.126  & 0.53 & 0.54 & 0.019 & \\ 
\noalign{\smallskip}
B80S1   &   &        & 0.624 & 0.074  & 0.09 & 0.76 & -0.001 & \\ 
B80D1   &   & DD2    & 0.706 & 0.115  & 0.21 & 2.13 & -0.002 & \\ 
S40S1   &   & SFHo   & 0.943 & 0.128  & 0.13 & 1.07 & $<$0.001 & \\ 
\end{tabular}
\end{table*}

\begin{figure*}   
\centering
\includegraphics[width=0.47\linewidth]{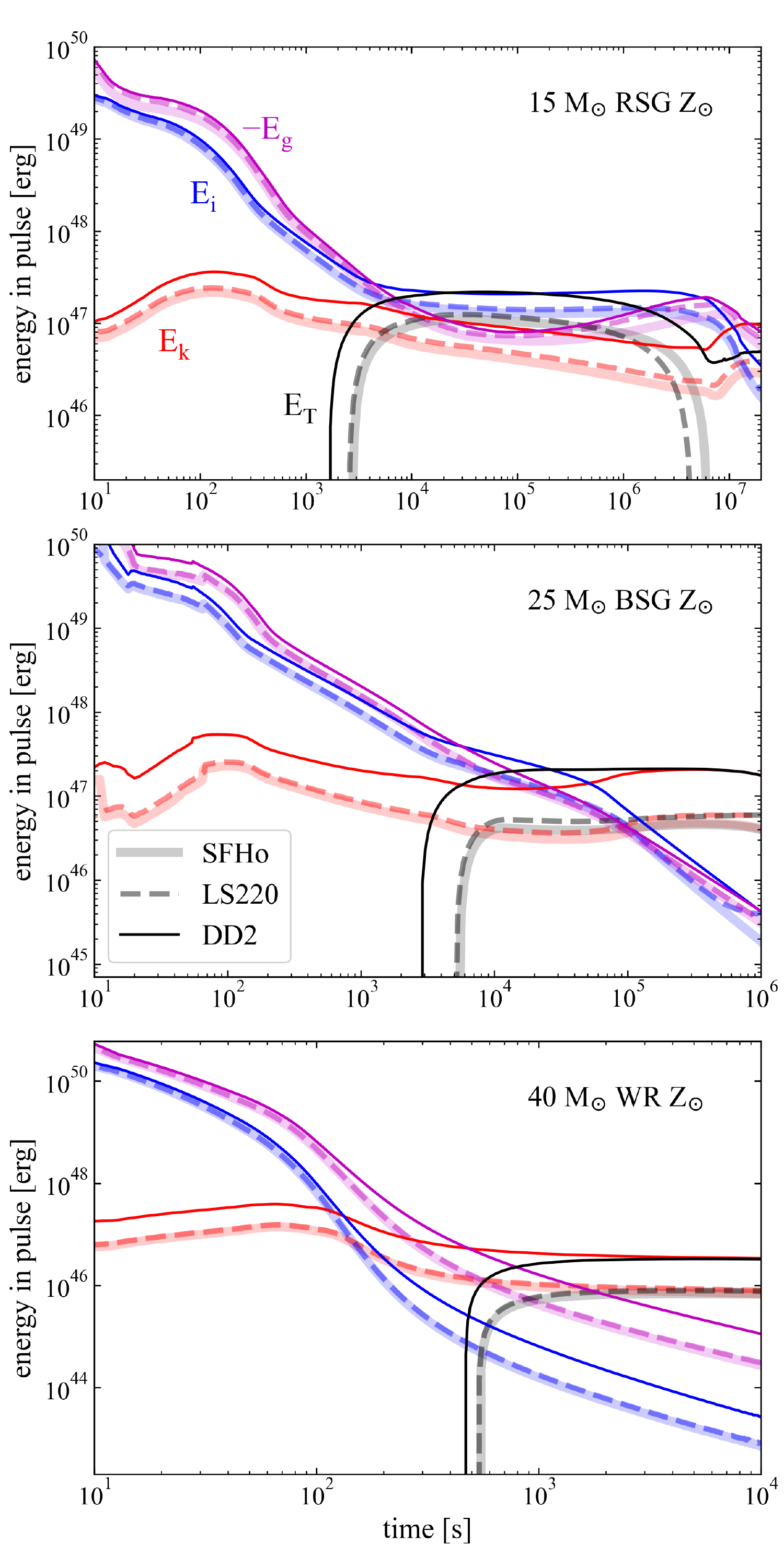}
\includegraphics[width=0.47\linewidth]{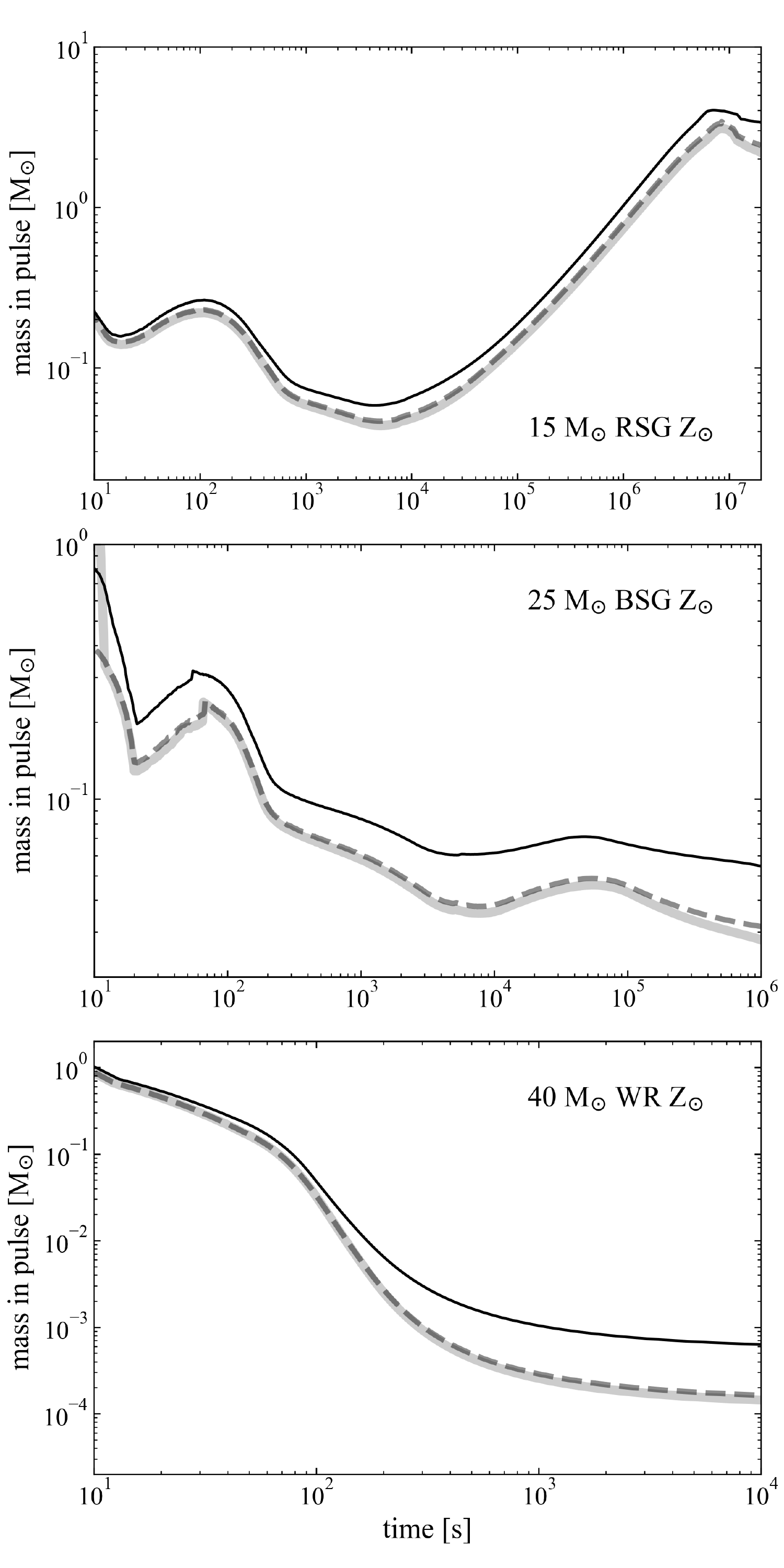}
\caption{\emph{Left:} Evolution of the energy components in the acoustic pulse generated by 
neutrino-induced gravitational mass decrease for our three fiducial progenitors: 
R15 (top), B25 (middle), and W40 (bottom). 
Each panel shows the kinetic (red), internal (blue), gravitational
(purple), and total energy (black) for models that interpolate $\Delta M_{\rm G}(t)$ from
{\tt GR1D}
and vary the EOS: DD2 (thin solid), SFHo (thick semitransparent), and LS220 (dashed).
Only positive total energies are shown.
\emph{Right:} Evolution of the mass contained by the pulse. Curves correspond
to the same models as in the left panel.
} 
\label{f:all_m4nr_E}
\end{figure*}

\subsection{Overview of evolution for different progenitors}
\label{s:comparing_progenitors}
\label{s:comparing_mloss}

The inner core evolution in all of our models is qualitatively the same. After
core-collapse and bounce, the shock stalls and then gradually recedes on a
timescale of $\sim$\,1\,s, as expected for non-rotating iron core progenitors
evolved in spherical symmetry
\citep{liebendoerfer01,rampp02,thompson03,sumiyoshi05}.  With the exception of
model R15D1, a {BH} forms within $3$\,s after bounce. At that
time, the difference between baryonic and gravitational masses $\Delta M_{\rm
G}$ stops increasing, as shown in Figure~\ref{f:dM_t}.

At a radius $\sim 10^9$\,cm, where the local free-fall time $t_{\rm ff}$ is
comparable to the time to change the gravitational mass via neutrino cooling
($\sim t_{\rm bh}$), an acoustic pulse forms which moves outward with a Mach number
of order unity (F18; \citealt{coughlin_2018a}). The subsequent evolution of
this sonic pulse and its effect on the stellar envelope depends on the type of
stellar progenitor. Table \ref{t:results} summarizes the results of our
hydrodynamic simulations.

For models in which we initialize the domain in {\tt FLASH} by remapping fluid 
quantities from {\tt GR1D} at $t=t_{\rm stall}$ (evolution mode 3 in Table~\ref{t:evolution_modes}), 
there are discrepancies at the $\lesssim 10\%$ in the fluid quantities 
(slightly lower density, lower internal energy, differing infall velocity) 
within the outermost mapping radius ($\sim 10^9$\,cm) relative to the presupernova model.
This results in the remapped models undergoing a slighly faster collapse than corresponding
models that only make use of $\Delta M_{\rm G}(t)$ to modify gravity (evolution modes 1 and 2),
which yields either a reduced kinetic energy of the shock and 
lower amount of mass ejected (models R15S3 and B25S3) or no mass ejected
at all (model W40S3).
Resolving this discrepancy requires a self-consistent treatment of the
entire star using a general-relativistic framework, which 
is beyond the scope of our study. In the rest of the paper we limit
the discussion to models that only interpolate $\Delta M_{\rm G}(t)$ from {\tt GR1D} or otherwise
approximate it analytically.

Figure~\ref{f:all_m4nr_E} shows the evolution of the energy in the sound
pulse as it travels outward through the envelope and becomes a running shock, 
for our three fiducial progenitors. Following F18, we define the pulse shell as 
the region limited by the forward edge of the pulse/shock on its front,
and by the stagnation point on its rear, which separates expanding from
accreting flow. Figure~\ref{f:all_m4nr_E} also shows that the mass in this pulse 
changes with time, as it 
sweeps up material on its front and loses it to fallback accretion
from its rear. In all cases, the kinetic energy in the pulse
is initially a small fraction of the gravitational and internal energies in the shell. 
As the pulse propagates out, the kinetic energy eventually becomes comparable 
and/or exceeds the (more rapidly decreasing) thermal and gravitational energies as it emerges from the 
stellar surface. The final net energy of the outgoing shock is comparable to its
initial kinetic energy. 

The propagation of weak shocks in gravitationally bound
stellar envelopes does not conserve energy \citep{coughlin_2018b}. Depending
on the radial dependence of the stellar density profile and on the 
initial strength of the pulse, the shock can accelerate or decelerate,
yielding behavior that ranges from a strong shock as it reaches the stellar surface to
fizzling as a rarefaction wave that ejects a negligible amount of mass 
(\citealt{coughlin_2019a,ro_2019}; see also \citealt{linial_2020}).

The energy of the acoustic pulse can be understood to order of magnitude from the
impulse $\Delta v_{\rm pulse}$ imparted to a mass shell by the change in
gravity over a free-fall time $t_{\rm ff}$, $\Delta v_{\rm pulse}(r) = (G\Delta
M_{\rm G}/r^2)\, t_{\rm ff}$ (\citealt{coughlin_2018a}, F18). The kinetic
energy in the pulse is 
\begin{equation}
\label{eq:ke_pulse}
\Delta E_{\rm pulse} \simeq \frac{1}{2}M_{\rm pulse}\Delta v_{\rm pulse}^2,
\end{equation}
with $M_{\rm pulse}\simeq 4\pi r^2 H_{\rm p}$, and $H_{\rm p}$ being the pressure scale
height. In terms of stellar quantities, we can write the maximum kinetic energy
that a stellar shell can have as (F18)
\begin{eqnarray}
\label{eq:epulse_analytic}
E_{\rm k,\, max} \simeq && 2.5\times 10^{47} \left(\frac{\alpha}{0.4}\right)\left(\frac{H_{\rm p}/r}{0.4}\right)
                     \left(\frac{\Delta M_{\rm G}}{0.15M_\odot} \right)^2\nonumber\\
         && \qquad\qquad \times \left(\frac{2\times 10^9\,\textrm{cm}}{r} \right)\,\textrm{erg}~,
\end{eqnarray}
where we have evaluated equation~(\ref{eq:ke_pulse}) at the location where energy extraction is maximal 
(point of acoustic pulse formation), and where $\alpha = d\ln M(r)/d\ln r$. 
The maximum kinetic energies obtained in the simulations (Figure~\ref{f:all_m4nr_E}, also shown in Table~\ref{t:results} 
as $E^{\rm sim}_{\rm k,max}$) are broadly consistent with 
equation~(\ref{eq:epulse_analytic}) given the characteristic gravitational mass changes $\Delta M_{\rm G}(t_{\rm bh})$
shown in Table~\ref{t:results}. 

The mass ejected via this mechanism is set, to order of magnitude, by the exterior mass
coordinate in the star at which the gravitational binding energy is comparable to the
shock energy. RSGs, with weakly bound hydrogen envelopes, can eject several solar masses
of slowly-moving material \citep{Lovegrove_Woosley_2013}, whereas BSGs and WR stars eject 
much smaller masses at higher speeds (F18).

\begin{figure}
\centering
\includegraphics[width=\linewidth]{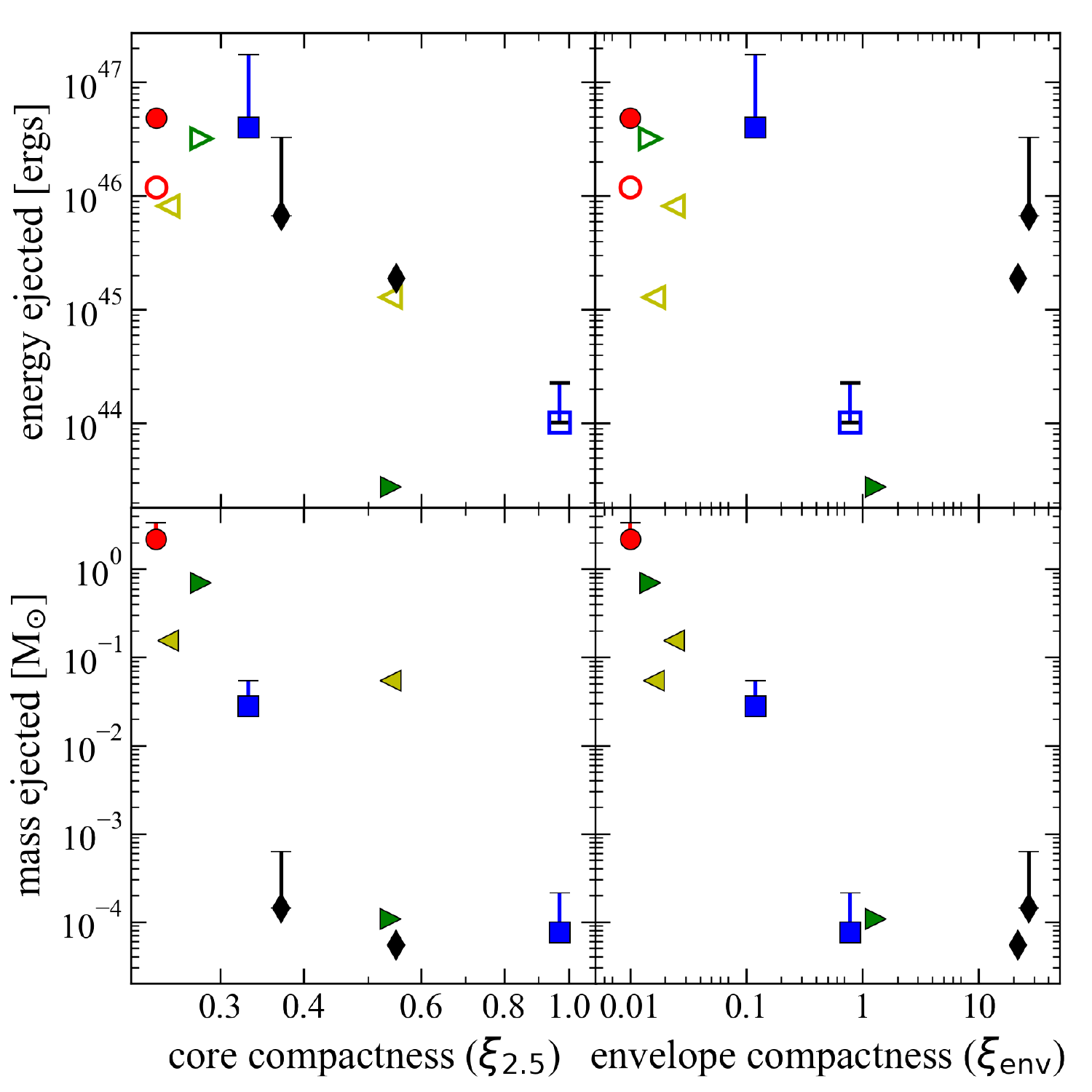}
\caption{Ejecta energy (top) and mass (bottom) as a function of core compactness 
$\xi_{2.5}$ (left; eq.~\ref{eq:compactness}) and envelope compactness 
$\xi_{\rm env}$ (right; eq.~\ref{eq:global_compactness}) for all models 
that interpolate $\Delta M_{\rm G}(t)$ from {\tt GR1D} using the SFHo EOS. The error 
bars indicate the change introduced, for our 3 fiducial progenitors (R15, B25, W40), 
by evolving the inner core with the DD2 EOS instead. Colored markers
correspond to their respective progenitors: red circles for RSGs, blue squares for 
BSGs, black diamonds for WRs, yellow left triangles for YSGs, green right triangles
for S20 and S40 models. Open and full symbols in the top row denote bound and unbound ejecta, 
respectively, with the exception of the R15 models, for which we have plotted both 
R15S1 (open circle, bound) and R15D1 (full circle, unbound).
}
\label{f:compactness_vs_M_ej+E_ej}
\end{figure}

\begin{figure*}
\includegraphics*[width=0.5\textwidth]{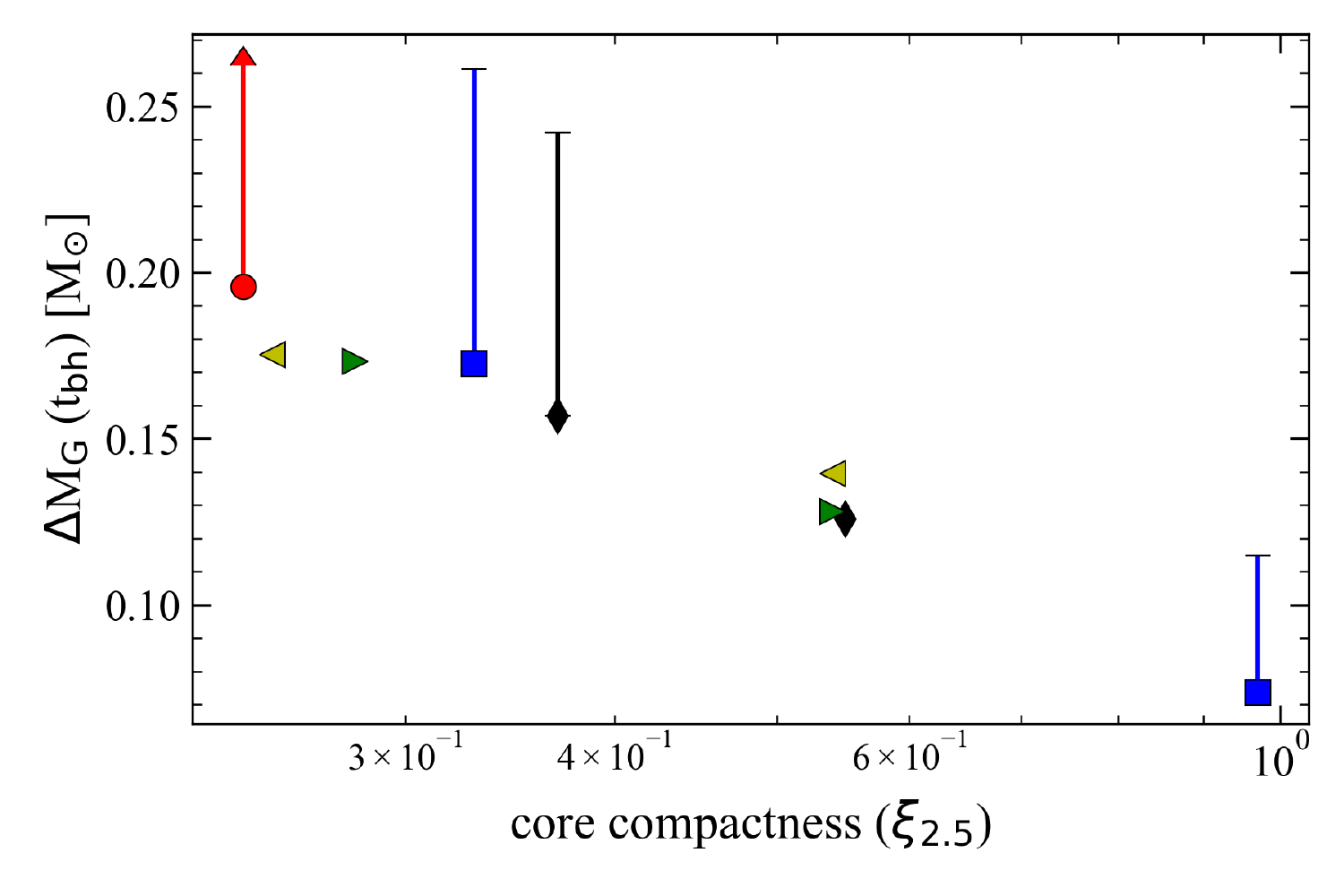}
\includegraphics*[width=0.5\textwidth]{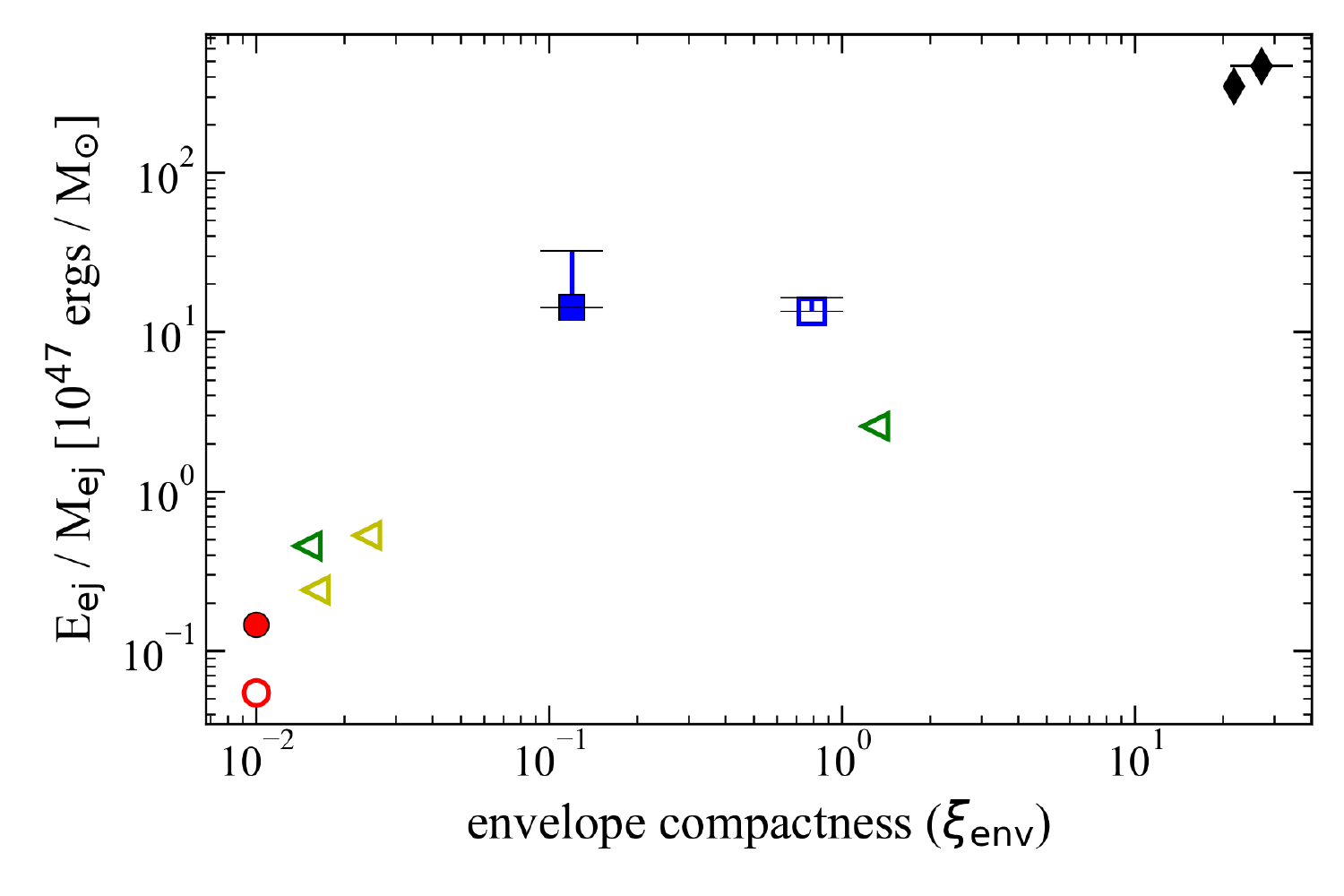}
\caption{\emph{Left:} Maximum gravitational mass lost to neutrino emission
$\Delta M_{\rm G}(t_{\rm bh})$ as a function of core compactness $\xi_{2.5}$.
\emph{Right}: Ejecta energy per unit mass as a function of envelope compactness
$\xi_{\rm env}$. Symbols and error bars have the same meaning as in
Figure~\ref{f:compactness_vs_M_ej+E_ej}, with the extra high-compactness BSG on the left panel 
representing model B80S1, which does not eject mass.  An upper limit is used for $\Delta
M_{\rm G}(t_{\rm bh})$ from model R15D1 since it did not collapse to a BH in
{\tt GR1D} within $4.2$\,s of evolution. 
}
\label{f:dM-v2_compactness}
\end{figure*}

Figure~\ref{f:compactness_vs_M_ej+E_ej} shows the ejecta mass and energies
as a function of core compactness (eq.~\ref{eq:compactness}) and 
envelope {compactness} (eq.~\ref{eq:global_compactness}), for all of our 
simulations that interpolate $\Delta M_{\rm G}(t)$ from {\tt GR1D} runs that use
the SFHo EOS.
The inverse relation between ejected mass and envelope compactness is evident:
given shock energies of characteristic magnitude of $\sim 10^{47}$\,erg, the
mass ejected is inversely proportional to the surface gravity of the star.
The inverse dependence of the ejecta energy with core compactness can be
understood from the monotonic decrease in
$\Delta M_{\rm G}(t_{\rm bh})$ with core compactness (Figure~\ref{f:dM-v2_compactness}). 
Higher compactness is associated with a shorter time to BH formation \citep{oconnor_2011,dasilva_2020}, 
which results in less gravitational mass lost to neutrinos.
While the ejecta energy is weakly correlated with envelope compactness, the 
energy per unit mass (and hence velocity) of the ejecta is strongly dependent on the envelope 
compactness (Figure~\ref{f:dM-v2_compactness}): stars that can unbind matter from their 
surfaces do so at speeds that scale with the escape velocity at that location
($v_{\rm esc}\propto \xi_{\rm env}^{1/2}$). 

To illustrate these trends with specific cases, we can compare models R15S1 and Y22S1, which have
similar envelope compactness (0.010 and 0.016 respectively), but the YSG has
more than double the core compactness of the RSG (0.54 versus 0.24,
respectively). Model R15S1 ejects about 2.2\,\Msun, a factor of 40
more than the YSG ejecta (5.39$\times 10^{-2}$\Msun), which follows from the
larger value of $\Delta M_{\rm G}(t_{\rm bh})$ for the RSG. In both cases, the
ejecta is bound (see \S\ref{s:comparing_eos} for a discussion of hydrogen recombination energy).
Likewise, models R15S1 and Y25S1 have similar core compactness (0.24 and 0.25,
respectively), but the YSG's envelope compactness is more than double that of
the RSG (0.010 and 0.024, respectively). The Y25 model ejects $0.155$\Msun,
which is approximately 14 times less compared to the $\sim$ 2.2\Msun~from the RSG. In
both cases the ejecta is bound, with the energy per unit mass being
larger in the YSG. Model S20S1 has higher core and envelope compactness than model R15S1, and 
also ejects bound mass. 

Model B80S1 has a higher core (0.97) and envelope compactness (0.79) than model B25S1 
(0.33 and 0.12 respectively). While B25S1 ejects 2.8$\times 10^{-2}$\,\Msun
with energy $4\times 10^{46}$\,erg by the end of the simulation, model B80S1
generates a shell with mass $\sim 8\times 10^{-5}$\,\Msun\, that is bound 
($E_{\rm ej}$\,$\sim$\,$-10^{44}$\,erg) by the time it approaches the stellar surface
and reaches the floor of temperature (after which time the internal energy starts rising
and the subsequent evolution is unreliable). This $80$\,\Msun\, progenitor also fails when using
parameterized neutrino mass loss in F18. Model S40S1, in contrast, is a BSG with similar
envelope compactness as B80S1 but smaller core compactness (0.54), ejecting marginally bound
mass in amounts comparable to the WR models, which have much higher envelope compactnesses.

Model W50S1 has a higher core compactness than model W40S1 (0.55 and 0.37 respectively), 
but a lower envelope compactness (22 and 27, respectively). Model W50S1 ejects less mass and 
with lower energy (5.4$\times 10^{-5}$\,\Msun\, and 1.9$\times 10^{45}$\,erg) than 
W40S1 (1.4$\times 10^{-4}$\,\Msun and 6.7$\times 10^{45}$\,erg). While the energy per unit
mass is comparable, it is higher in the progenitor with higher envelope compactness (W40S1).

A systematic uncertainty of $\sim 10\%$ in our results using the default
evolution mode arises from our choice of $r_{\rm in}=2\times 10^8$\,cm.
Model W40S1.7 is identical to W40S1 except for the
position of the inner radial boundary set at $r_{\rm in}=2\times 10^7$\,cm. The
resulting ejecta mass and energy are about $30\%$ higher in the model
with the smaller inner radial boundary (see also F18 for further tests
on the sensitivity of the choice of boundary position).

\subsection{EOS Dependence}
\label{s:comparing_eos}

Table~\ref{t:results} shows that for the same stellar model and spatial
resolution, using the stiffer DD2 EOS to evolve the inner core results in
more mass ejected and with higher specific energy, by a factor of several,
relative to using the softer SFHo or LS220 EOS. Equation~(\ref{eq:epulse_analytic})
shows that the maximum kinetic energy of the shock is proportional
to the square of the gravitational mass lost to neutrinos $\Delta M_{\rm G}$. 
All models evolved with the DD2 EOS achieve higher values of $\Delta M_{\rm G}(t_{\rm bh})$
than their equivalent models evolved with the SFHo EOS by a factor of up to
two. Results obtained with the SFHo and LS220 EOSs are usually very similar to each 
other and often overlap. 

The origin of this trend with EOS stiffness is illustrated in
Figure~\ref{f:dM_t}: the stiffer DD2 EOS yields a longer time to BH formation
and therefore results in more gravitational mass lost to neutrino emission than
in the models that use the SFHo EOS. While there are some differences in the
growth rate of $\Delta M_{\rm G}$, associated {with} the differing neutrino
luminosities, which in turn are most sensitive to the effective nucleon masses in the
EOS \citep{schneider_2019}, these luminosity differences are sub-dominant
compared to those arising from the time interval during which the PNS emits
neutrinos, which is set primarily by the accretion rate and the maximum mass 
{at finite entropy} ($M_{\rm tov}$) 
that the EOS can support{, which depends on both cold and thermal pressure
components} \citep{hempel_2012,dasilva_2020}. The
same trend of increasing ejecta mass and energy with increasing $M_{\rm tov}$
was found by \citet{Lovegrove_Woosley_2013} using a parameterized evolution of
the inner core.

Figure~\ref{f:compactness_vs_M_ej+E_ej} shows the spread in ejected masses due
to the EOS (represented as error bars) for our three fiducial progenitors.
Aside from the magnitude of the gravitational mass lost and a minor change in
the location of the acoustic pulse formation (radius at which  $t_{\rm bh} \simeq
t_{\rm ff}$), the evolution of the shock as it propagates into the envelope is
qualitatively the same for all EOSs, as shown in Figure~\ref{f:all_m4nr_E}

A key quantitative difference introduced by the EOS is that when using SFHo or LS220, none
of the RSG progenitors eject unbound mass (Table~\ref{t:results}). The equation
of state used in {\tt FLASH} assumes fully ionized nuclei, but it is known that
hydrogen recombination can dominate the energetics upon subsequent expansion of
the shock in failed supernovae from RSGs \citep{Lovegrove_Woosley_2013}. To
estimate the importance of this missing effect, we compute the recombination
energy $E_{\rm rec}$ that can be liberated if all the hydrogen contained in the
ejecta recombines {to the ground state}, 
\begin{eqnarray}
\label{eq:Erec_definition}
E_{\rm rec} & = & \frac{M_{\rm H}}{m_{\rm p}} \chi_{\rm H}\\
M_{\rm H}   & = & \int_0^{M_{\rm ej}} X(M) dM,
\end{eqnarray}
with $X$ the mass fraction of hydrogen in the ejecta, $m_{\rm p}$ the proton mass, and
$\chi_{\rm H}=13.6$\,eV. 
The resulting recombination energies for RSGs and YSGs are shown in
Table~\ref{t:results}.  With the exception of model s20 and the model that maps
initial profiles from {\tt GR1D} (R15S3), the ejecta for all other RSG progenitors
should be unbound after including this contribution.  Regarding YSGs, model Y22S1 can
potentially unbind its ejecta marginally when including hydrogen recombination (e.g.,
with a more precise calculation than our rough estimate), while for model Y25S1 
recombination energy falls short by a factor 3 and unbinding is unlikely.

\begin{table}
\caption{Comparison of the ejecta masses and energies obtained in F18
(high-resolution ``e\_HR" models, using parametric inner core evolution)
and in this work (interpolating $\Delta
M_{\rm G}(t)$ from {\tt GR1D} using either the SFHo [S1] or DD2 [D1] EOS
and at the same resolution as F18; c.f.
Table~\ref{t:resolution}), for our three fiducial progenitors
(Table~\ref{t:progenitors}).}
\centering
\label{t:f18_comparison}
\begin{tabular}{l|cc|cc}
\noalign{\smallskip}
Progenitor   & \multicolumn{2}{c}{$M_{\rm ej} (M_\odot)$} & \multicolumn{2}{c}{$E_{\rm ej} (10^{47}$\,erg)}\\
      & [S1,D1]                & F18                 & [S1,D1]     & F18     \\
\hline                                                               
R15   & [2.2, 3.4]             & 4.2                 & [-0.1, 0.5] & 1.9     \\
B25   & [0.03, 0.05]           & 0.05                & [0.5, 1.8]  & 1.6     \\
W40   & [1, 6]$\times 10^{-4}$ & $5\times 10^{-4}$   & [0.06,0.31] & 0.25    \\
\end{tabular}
\end{table}

The massive BSG progenitor B80 fails with both the SFHo and DD2 EOSs. While
the latter yields a larger shock kinetic energy by a factor $\sim 2$, the qualitative result does not
change: as the shock approaches the stellar surface, it reaches the floor
of temperature and the internal energy begins rising. Prior to that, the
mass in the shell is $\lesssim 10^{-4}$\,$M_\odot$ and the outward-moving material
is gravitationally bound with $E_{\rm ej}\simeq -10^{44}$\,erg. Further study of 
these failing models will require a low-temperature EOS consistent with that used
to evolve the presupernova model.

Comparing to the results of F18, which were obtained using an analytic model
for the inner core evolution and therefore of $\Delta M_{\rm G}(t)$, we find
that their results are in broad agreement with our models that use the DD2 EOS.
In contrast, results obtained with the softer SFHo EOSs have energies lower by
a factor of several compared to F18. A side-by-side comparison of results for
the same fiducial progenitors and at the same spatial resolution is shown in
Table~\ref{t:f18_comparison}.  The relation between the two sets of results
follows from the fact that F18 assumed $M_{\rm tov} = 2.5$\,$M_\odot$ as an input
{to} their parametric neutrino scheme. The maximum NS mass at an
entropy of $4k_{\rm B}$ per baryon correlates well with $t_{\rm bh}$ when
comparing among different EOSs \citep{hempel_2012}. The choice of F18 is much
closer to the finite-entropy $M_{\rm tov}$ for the DD2 EOS ($2.57$\,\Msun~at
$4k_{\rm B}$ per baryon; M. Hempel, private communication) than for the SFHo
EOS ($2.3$\,\Msun~at $4k_{\rm B}$ per baryon, \citealt{steiner_2013}).

\subsection{Simplified inner core evolution: linear ramp for $\Delta M_{\rm G}$}

\begin{figure}
\centering
\includegraphics[width=\linewidth]{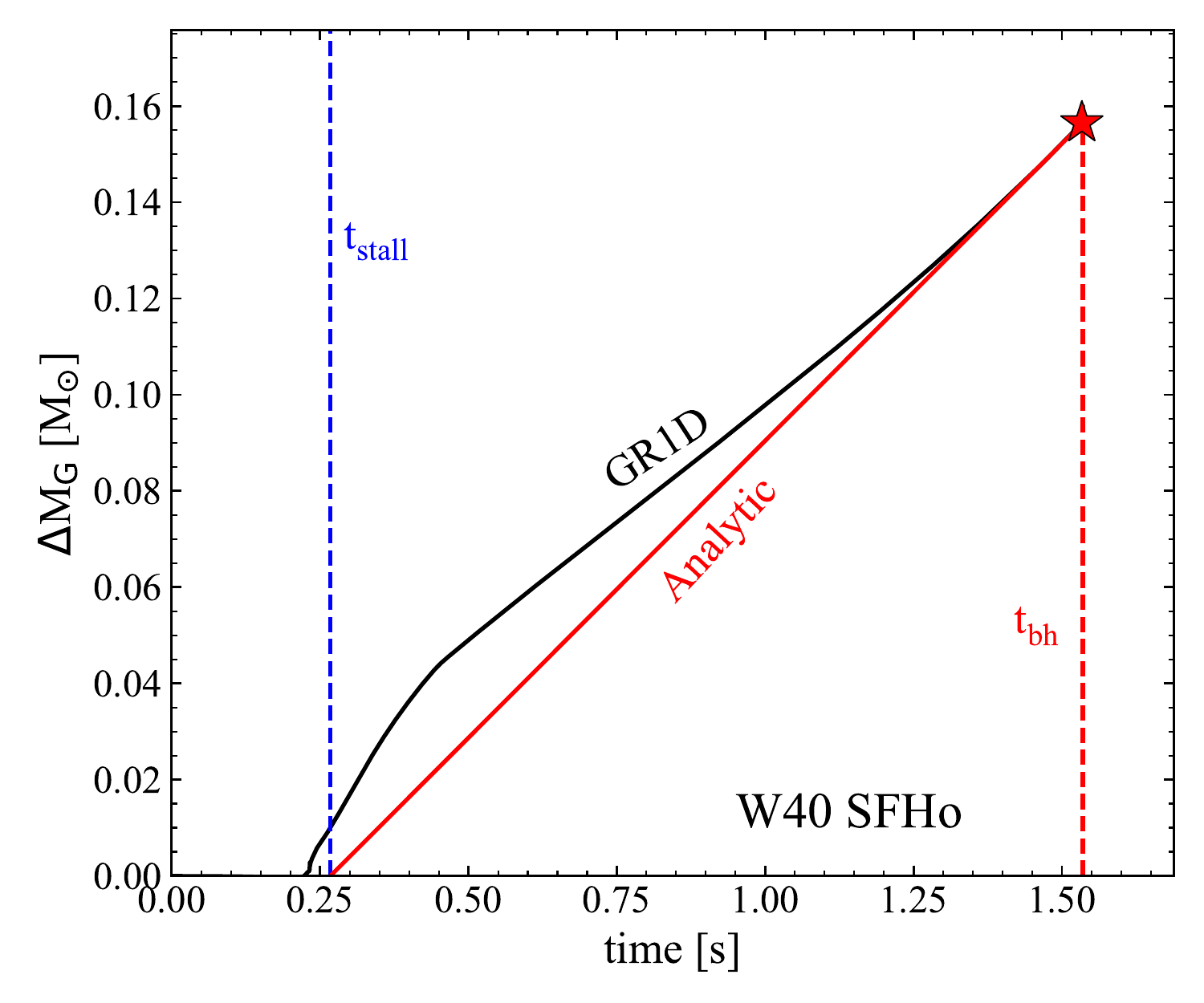}
\caption{Evolution of $\Delta M_{\rm G}$ as a function of time since core-collapse 
for the W40 progenitor. The result from {\tt GR1D} using the SFHo EOS is shown
in black, while our linear ramp parameterization is shown in red, with the
star denoting the time of BH formation. The vertical dashed blue line shows
the time $t_{\rm stall}$ at which the shock stalls, which we use to start the ramp function
(same time as in Figure~\ref{f:velx_csnd_rcut}). The vertical dashed red line 
shows the time of BH formation t$_{\rm bh}$.
}
\label{f:W40S_m4nr_vs_m5_dM}
\end{figure} 

To assess the sensitivity of mass ejection to the detailed history of neutrino
emission by the protoneutron star before BH formation, we explore a set of
models in which we parameterize $\Delta M_{\rm G}(t)$ as a linear ramp in time
(Figure~\ref{f:W40S_m4nr_vs_m5_dM}). The input parameters are the maximum value
of $\Delta M_{\rm G}$, the time of BH formation $t_{\rm bh}$, and a starting
time, which we choose to be that when the shock radius reaches its maximum value
($t_{\rm stall}$, same as that used when remapping the domain from {\tt GR1D}).
These parameters are generally reported in (or otherwise are straightforward to
obtain from) published studies of BH formation in failed SNe.

\begin{figure}
\centering
\includegraphics[width=\linewidth]{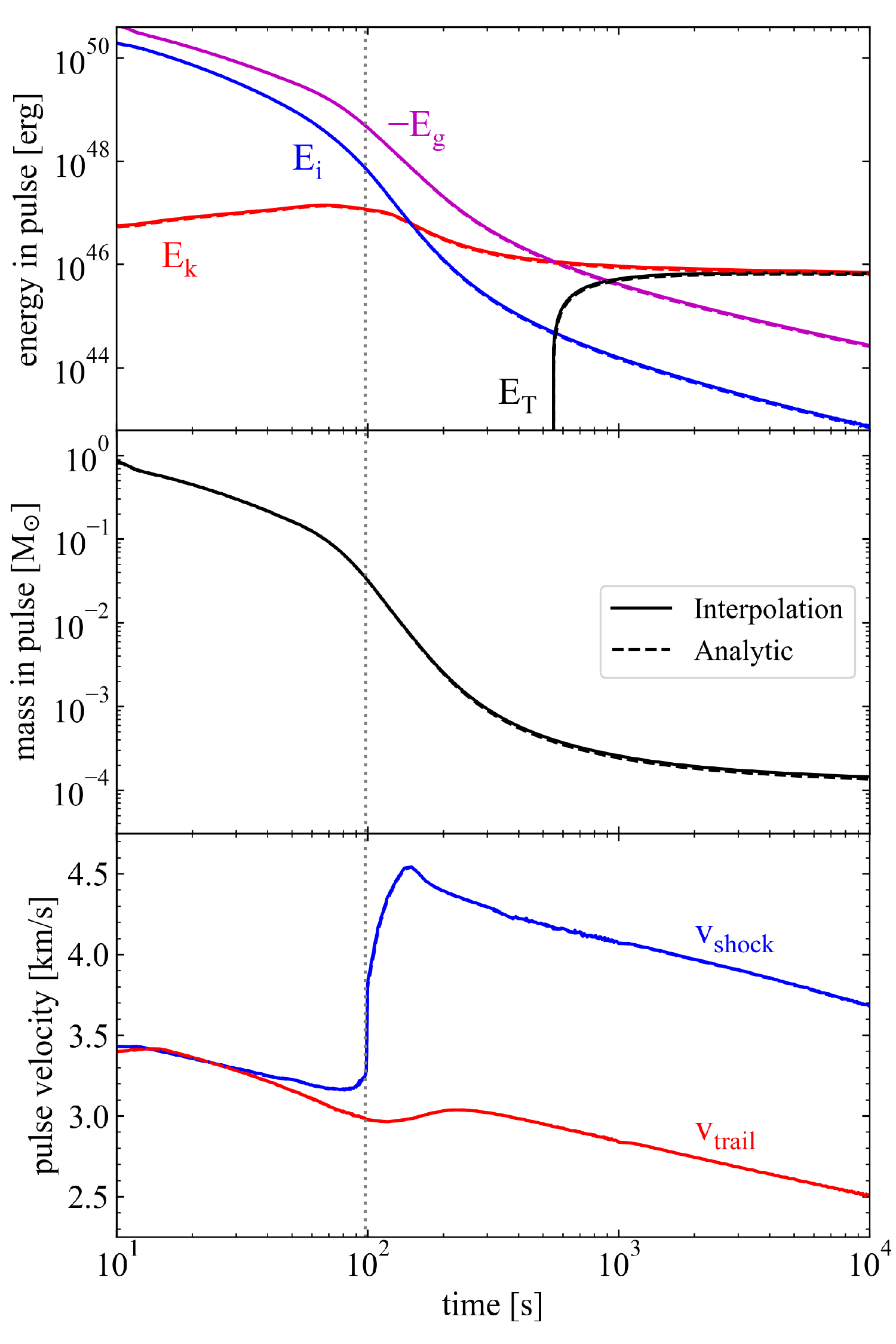}
\caption{Comparison between models W40S1 and W40S2, which interpolate $\Delta M_{\rm G}(t)$ 
from {\tt GR1D} or use an analytic ramp model (Figure~\ref{f:W40S_m4nr_vs_m5_dM}), respectively. 
Top, middle, and bottom panels show the evolution of the energies, mass, and velocity of the 
sound pulse, as labeled ($v_{\rm shock}$ and $v_{\rm trail}$ are the velocities of the forward
and rear end of the shell, respectively). The vertical dotted line indicates the time at which the shock 
emerges from the stellar surface.
}
\label{f:W40S_m4nr_vs_m5_EMv}
\end{figure}

Figure~\ref{f:W40S_m4nr_vs_m5_EMv} compares the evolution of the energy, mass,
and velocities of the outgoing shell for models W40S1 and W40S2, with the former
interpolating $\Delta M_{\rm G}(t)$ from {\tt GR1D} and the latter using the
analytic ramp prescription.  The shock properties are very close to one another
in both models, with a relative difference of $6\%$ in energy and in
mass by the end of the simulation.  A smaller relative difference in final
energy and mass (few percent) is found for the corresponding BSG models (B25S1-B25S2),
while for the RSG the mass ejected is nearly the same whereas the
negative ejecta energy differs by 30\% (Table~\ref{t:results}).

The low sensitivity to the detailed neutrino history can be understood from the
fact that variations in neutrino emission (as inferred from
Figure~\ref{f:dM_t}) occur on timescales much shorter than $t_{\rm bh}$. Mass
shells for which the local dynamical time $t_{\rm dyn}$ is comparable to this
neutrino variability timescale are accreted into the BH before there is
sufficient time to affect the emergence of the sound pulse at  a location such
that $t_{\rm dyn}\sim t_{\rm bh}$. Marginally bound shocks (from RSGs) are more
sensitive to small changes in neutrino emission history than those from progenitors
that eject mass more robustly.

\subsection{Effect of spatial resolution}
\label{s:resolution}

\begin{table*}
\caption{Resolution dependence of key quantities for selected models.  Columns
from left to right show model name, EOS used in {\tt GR1D}, spatial resolution
in the {\tt FLASH} run
($\Delta r / r = \{4.5,1.1,0.6\}\times 10^{-3}$ corresponds to
$\{512,2048,4096\}$ cells per decade in radius in a logarithmic grid), mass
ejected in the {\tt FLASH} run $M_{\rm ej}$, total energy of ejecta in the {\tt
FLASH} run $E_{\rm ej}$, shock breakout velocity measured in the {\tt FLASH}
simulation, and analytic breakout velocity (eq.~[\ref{eq:vbo_wk}]).
}
\centering
\label{t:resolution}
\begin{tabular}{lcccccc}
\hline
\noalign{\smallskip}
Model & EOS & $\Delta r/r$ &  $M_{\text{ej}}$  & $E_{\text{ej}}$  & $v^{\rm num}_{\rm bo}$ & $v^{\rm WK}_{\rm bo}$\\
      &     & ($10^{-3}$)  &  $(\Msun)$        & ($10^{47}$\,erg) & (km\,s$^{-1}$) & (km\,s$^{-1}$) \\
\noalign{\smallskip}
\hline
R15S1 & SFHo & 4.5  & 2.20 & -0.130 & 60 & ... \\
R15S1 &      & 1.1  & 2.19 & -0.119 & 60 & ... \\
R15D1 & DD2  & 4.5  & 3.38 &  0.498 & 80 & 30  \\
R15D1 &      & 1.1  & 3.37 &  0.489 & 80 & 30  \\
\noalign{\smallskip}
\hline
\noalign{\smallskip}
      &      &  & $M_{\text{ej}}$    & \\
      &      &  & ($10^{-2}$\,\Msun) & \\
\noalign{\smallskip}
\hline
\noalign{\smallskip}
B25S1 & SFHo & 4.5  & 2.90  & 0.491 &  900  & 600 \\ 
B25S1 &      & 1.1  & 2.80  & 0.399 & 1,000 & 500 \\ 
B25D1 & DD2  & 4.5  & 5.46  & 1.77  & 1,300 & 900 \\ 
B25D1 &      & 1.1  & 5.45  & 1.76  & 1,300 & 900 \\
\noalign{\smallskip}
\hline
\noalign{\smallskip}
        &   &  & $M_{\text{ej}}$    & \\
        &   &  & ($10^{-4}$\,\Msun) & \\
\noalign{\smallskip}
\hline
\noalign{\smallskip}
W40S1   & SFHo & 4.5  & 1.32 & 0.059 &  9,000 & 9,000  \\
W40S1   &      & 0.6  & 1.44 & 0.067 & 17,000 & 10,000  \\
W40D1   & DD2  & 4.5  & 6.15 & 0.308 & 10,000 & 13,000 \\
W40D1   &      & 0.6  & 6.30 & 0.326 & 20,000 & 13,000 \\
\noalign{\smallskip}
\end{tabular}
\end{table*}

Table~\ref{t:results} reports the mass ejected and final shock energy for
selected models at two spatial resolutions: the same as in F18 ($\Delta r/r =
4.5\times 10^{-3}$) for comparison, and at 4-8 times finer grid spacing per
decade in radius  ($\Delta r/r = 0.6-1.1\times 10^{-3}$). The highest
resolution is used to resolve the surface pressure scale height of the W40
model with $\sim 3$ cells ($H_{\rm p}/R_{\rm cc}\simeq 1.5\times 10^{-3}$).
For RSGs and BSGs, the surface pressure scale height is resolved with $\{2-10,4-20\}$
cells in our low-high resolution models, respectively.

The ejecta masses and energies are essentially converged with resolution 
for the RSG and BSG models that use the DD2 EOS (R15D1 and B25D1). 
Their SFHo counterparts (R15S1 and B25S1), with weaker explosions, are 
somewhat sensitive to resolution, with changes of a few percent in their 
ejected mass and $10-20\%$ in ejecta energies when increasing the resolution
by a factor 4. For the WR progenitor, on the other hand,
resolving the surface pressure scale height even with a few cells results in an 
increase of $10\%$ in ejecta mass and energy for model W40S1, with a
$2-5\%$ change in the more energetic model W40D1.
We expect further increases in resolution to behave in the
same way as for RSG and BSG models. As discussed in \S\ref{s:comparing_progenitors},
an additional $\sim 10\%$ uncertainty in ejecta mass and energy arises from the 
position of the inner radial boundary.

As an additional diagnostic of our simulations, we examine the velocity
at shock breakout, and compare with analytic formulae (as reviewed in
\citealt{waxman_2017}). The latter predict a shock breakout velocity
\begin{equation}
\label{eq:vbo_wk}
v_{\rm bo}^{\rm WK}\simeq
v_* \times 
\begin{dcases}
13\,M_{\rm ej,10}^{0.16}\,v^{0.16}_{*,8.5}\,R^{-0.32}_{\rm cc,12}  & \qquad (\textrm{BSG, WR})\\
\noalign{\smallskip}
4.5\,M_{\rm ej,10}^{0.13}\,v^{0.13}_{*,8.5}\,R^{-0.26}_{\rm cc,12} & \qquad (\textrm{RSG}),
\end{dcases}
\end{equation}
where $v_* = \sqrt{E_{\rm ej}/M_{\rm ej}}$, $M_{\rm ej,10}=M_{\rm }/(10\Msun)$, 
$v_{*,8.5}=v_*/(10^{8.5}\textrm{\,cm\,s}^{-1})$,
and $R_{\rm cc,12}=R_{\rm cc}/(10^{12}\textrm{\,cm})$, and where we have 
ignored the dependence on opacity and detailed stellar density profile dependence. 
The corresponding values of $v_{\rm bo}^{\rm WK}$ are shown in Table~\ref{t:resolution}
for selected models.

\begin{figure}
\includegraphics*[width=\columnwidth]{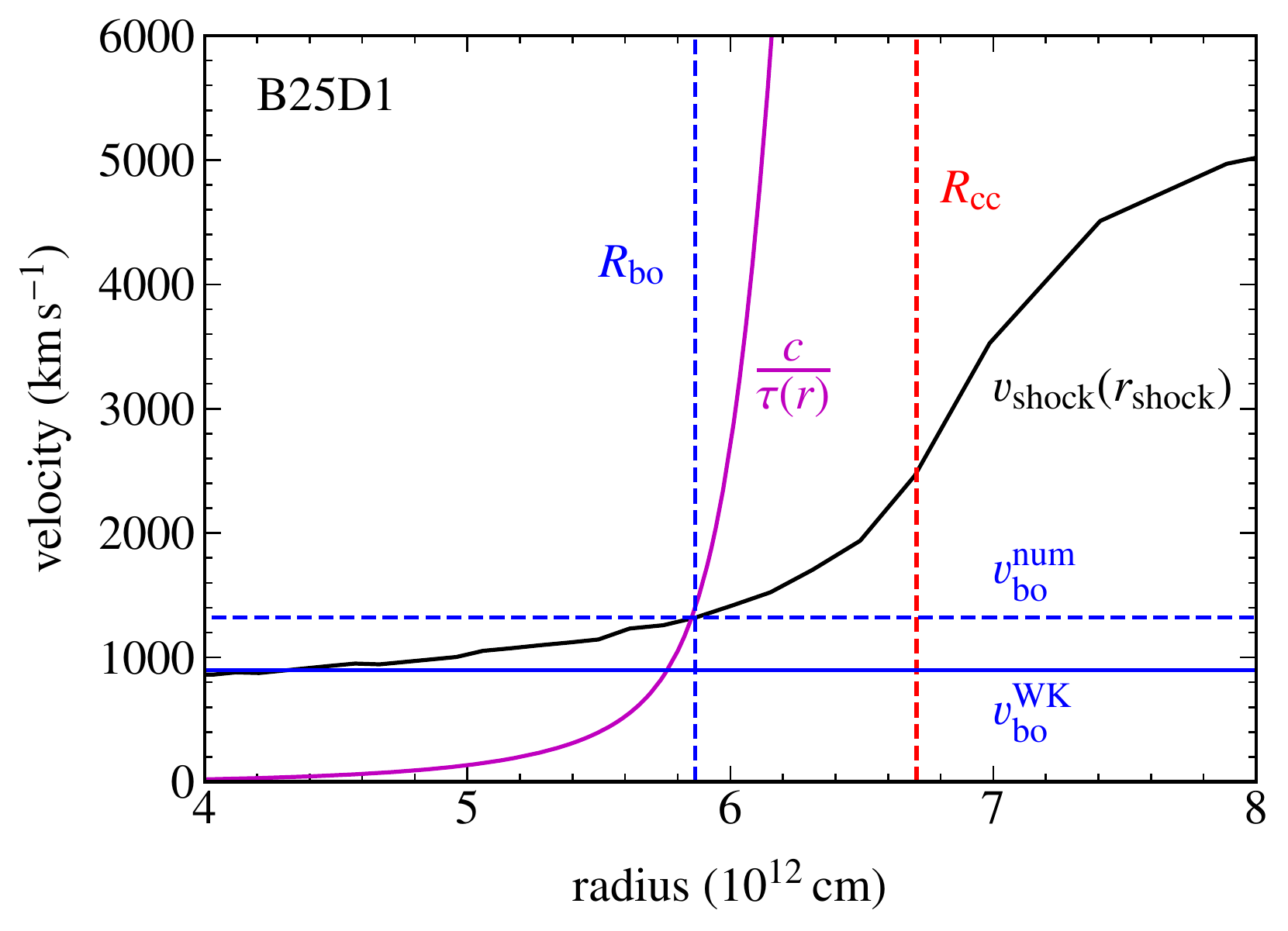}
\caption{Forward shock velocity $v_{\rm shock}$ as a function of forward shock
radius $r_{\rm shock}$ in model B25D1 at high resolution.  The numerical
shock breakout velocity $v_{\rm bo}^{\rm num}$ is obtained by searching for the
position at which $v_{\rm shock} = c/\tau(r)$, where $\tau(r)$ is the optical
depth in the presupernova star. For reference, we also show the analytic value
$v_{\rm bo}^{\rm WK}$ from equation~(\ref{eq:vbo_wk}).  The numerical shock
velocity is not smoothed for this model.}
\label{f:vbo_radius}
\end{figure}

The shock breakout velocity is measured from the {\tt FLASH} simulations as
follows. We use the position of the forward shock $r_{\rm shock}$ as a function of
time to calculate the forward shock velocity $v_{\rm shock}$ using time-centered
finite difference. When the resulting velocity curve is too  noisiy, we smooth this
function using a Savitsky-Golay filter to suppress fluctuations. We then obtain
the shock breakout velocity by an iterative process: we start with the shock
velocity at the stellar surface, and compute the shock breakout optical depth
$\tau_{\rm bo} = c/v_{\rm shock}$, which is then used to 
obtain the shock breakout radius $R_{\rm bo}$ using the optical depth profile 
in the presupernova progenitor. The shock velocity at the newly obtained position 
$r=R_{\rm bo}$ is then used to compute a new optical depth, which then yields another 
value for $R_{\rm bo}$. This process converges after a few iterations.

Table~\ref{t:resolution} shows the numerical shock breakout velocities for
selected models. Agreement with analytical values is  good to within a factor
of $2$. Figure~\ref{f:vbo_radius} shows an example of the
forward shock velocity profile and its relation to the analytic value for model
B25D1.  

\begin{figure*}
\includegraphics*[width=\textwidth]{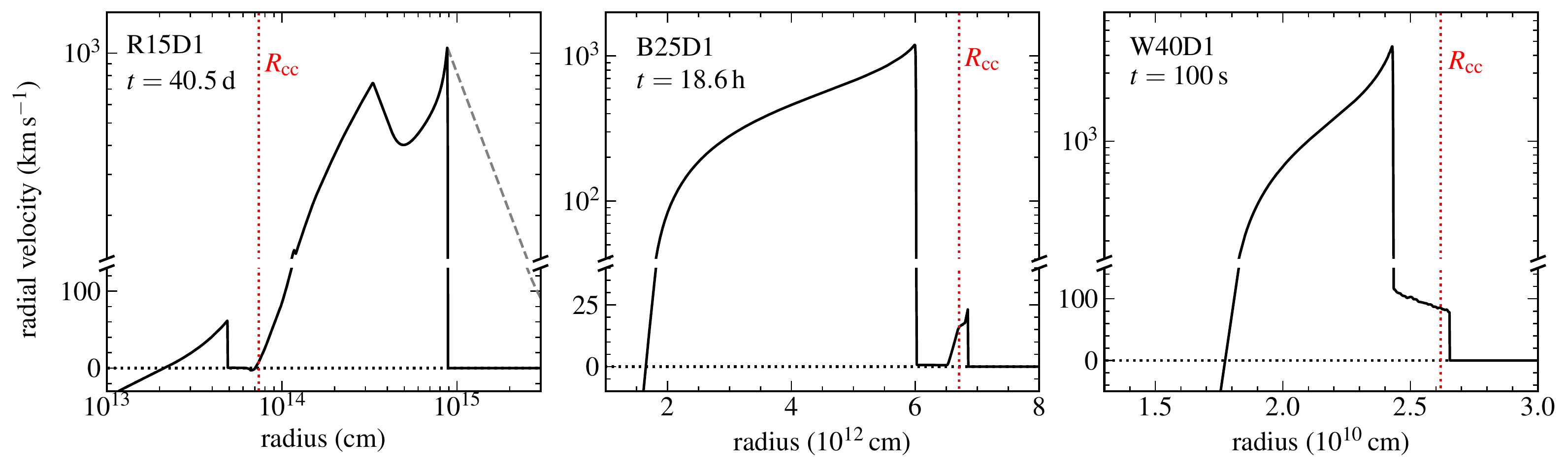}
\caption{Radial velocity profile of stellar matter at a time just before the 
main shock reaches the pre-collapse stellar surface (red vertical line labeled
$R_{\rm cc}$), for selected models at their highest resolutions. The gas near 
the photosphere acquires
positive velocity  even before the arrival of the shock, with noticeable
`surface' shocks in the RSG and BSG models. The dashed grey line shows the
velocity of ambient material in model R15D1 (the main shock has a smaller
velocity but carries more mass than the surface shock).}
\label{f:surface_shock}
\end{figure*}

Given that our simulations resolve the surface pressure scale height in all models, we
revisit the prediction of \citet{coughlin_2018a} regarding the outward acceleration
of the photosphere ahead of the main shock. A photospheric shock is possible 
given that the entire envelope experiences a nearly instantaneous change in the acceleration 
of gravity, and regions near the surface of the star have a rapidly decreasing sound speed with radius.
Figure~\ref{f:surface_shock} shows radial profiles of velocity at a time
when the main shock is about to reach the original pre-collapse surface at $r=R_{\rm cc}$.

The RSG model is such that outward acceleration of matter near the photosphere is
very efficient, launching a series of surface shocks that reach $\sim 10\,R_{\rm cc}$
by the time the main shock approaches the stellar surface. These surface shocks
contain very little mass, and include ambient material. Part of this dynamics
is likely to arise from the temperature being near the floor value of the EOS 
($10^4$\,K), which is higher than the actual surface temperature of the RSG and 
therefore implies inconsitent thermodynamics (fully ionized ions assumed in {\tt FLASH})
relative {to} the {EOS} used to construct the presupernova model.

A cleaner surface shock is visible in the BSG model, which displays a quiescent
ambient ahead of it. The magnitude of this surface shock is smaller by a
factor of almost $\sim 100$ compared to the main shock, and it does not
lead to a significant expansion of the stellar surface ahead of the arrival
of the main shock. For the WR model, the entire region inside the photosphere 
moves outward at $\sim 100$\,km\,s$^{-1}$ and, while no independent shock is visible 
before the arrival of the much faster main shock, this photospheric expansion is
significantly faster than in the RSG and BSG models, as predicted by  \citet{coughlin_2018a}.
Like in the BSG model, however, the expansion of the star is minimal before the arrival 
of the main shock. 

\subsection{Implications for electromagnetic counterparts}

\begin{table*}
\centering
\caption{Estimates for the bolometric shock breakout and plateau emission for our 
fiducial progenitors, which interpolate $\Delta M_{\rm G}(t)$ from {\tt GR1D} and use
the DD2 or SFHo EOS. Columns from left to right show: model name, shock breakout luminosity, 
breakout time (maximum between diffusion and light-crossing times),
shock velocity at breakout $v_{\rm bo}$, 
effective temperature at breakout $T_{\rm bo}$, plateau luminosity, plateau duration, and 
final shock velocity $v_{\rm exp} = \sqrt{2E_{\rm ej} / M_{\rm ej}}$. Model R15S1
ejects bound material so no plateau properties are computed, and the breakout velocity
is obtained from the simulation (Table~\ref{t:resolution}); for all other models the breakout velocity is computed
using equation~(\ref{eq:vbo_wk}).
For reference, the pre-supernova luminosities are
$\{1.3,3.8,5.7\}\times 10^5L_\odot$ for models R15, B25, and W40, respectively.
}
\label{t:observations}
\begin{tabular}{lccccccc}
\hline
\noalign{\smallskip}
Model & $L_{\rm bo}$      & $t_{\rm bo}$ & $v_{\rm bo}$         & $T_{\rm bo}$  & $L_{\rm pl}$     
  & $t_{\rm pl}$ & $v_{\rm exp}$ \\ 
      &   ($10^6$\,\Lsun) &              &     (km\,s$^{-1}$)   &   ($10^4$\,K) &  ($10^5$\,\Lsun) 
  &  (d)         &  (km\,s$^{-1}$) \\ 
\noalign{\smallskip}
\hline
\noalign{\smallskip}
R15S1       & $0.7$  & $4$\,d   & $60$*   & $0.5$ & ...  & ...   & ...   \\ 
R15D1       & $2$    & $8$\,d   & $40$    & $0.6$ & $2$  & $400$ & $40$  \\ 
\noalign{\smallskip}
B25S1       & $50$   & $6$\,h & $500$   & $5$   & $9$  & $20$  & $400$ \\    
B25D1       & $300$  & $3$\,h & $900$   & $8$   & $20$ & $20$  & $600$ \\   
\noalign{\smallskip}
W40S1       & $400$ & $1$\,s & $9,600$  & $140$ & $0.3$ & $1$ & $2,000$ \\  
W40D1       & $600$ & $1$\,s & $13,000$ & $150$ & $0.6$ & $2$ & $2,000$ \\  
\end{tabular}
\end{table*}

Non-rotating failed supernovae are expected to generate electromagnetic
emission in the form of shock breakout \citep{Piro_2013,lovegrove_2017} and
plateau emission \citep{Lovegrove_Woosley_2013}.
The predicted plateau properties for a RSG progenitor 
are consistent with those of the failed supernova candidate N6946-BH1
\citep{gerke_2015,adams_2017,basinger_2020}.  A wider range in 
emission properties is expected when accounting for WRs and BSGs (F18).

Following F18, we estimate bolometric shock breakout properties in our models using a 
combination of analytic formulate and quantities measured from the {\tt FLASH} simulations. 
The shock breakout luminosity is estimated as \citep{Piro_2013}
\begin{equation}
L_{\rm bo} \simeq \frac{E_{\rm rad}}{\max(t_{\rm lc},t_{\rm diff})},
\end{equation}
where $t_{\rm lc} = R_{\rm cc}/c$ is the light-crossing time over the 
stellar radius and $t_{\rm diff} = (R_{\rm cc}-R_{\rm bo})/v_{\rm bo}$ is
the radiation diffusion (and shock crossing) time, with $R_{\rm bo}$ the stellar radius 
at which the optical depth is $c/v_{\rm bo}$. The radiation energy $E_{\rm rad}$
is obtained for WRs using the formulae in \citet{waxman_2017}, while for BSGs and RSGs
it is measured directly from the simulation ($E_{\rm rad} = \int a T^4 dV$ over
the {volume between $R_{\rm bo}$ and $R_{\rm cc}$} 
at the time when {$r_{\rm shock}=R_{\rm cc}$}). The shock breakout
velocity is set to the analytic value in equation~(\ref{eq:vbo_wk}) except
for model R15S1, which has $E_{\rm ej} <0$, in which case we use the
value obtained from the simulation (Table~\ref{t:resolution}). The
temperature of the breakout emission is estimated using the luminosity
and stellar radius, $L_{\rm bo} = 4\pi R_{\rm cc}^2\,\sigma T_{\rm bo}^4$.

Table~\ref{t:observations} shows the shock breakout emission estimates for our
fiducial models employing the DD2 or SFHo EOSs. For reference, the
pre-supernova luminosities are $\{1.3,3.8,5.7\}\times 10^5L_\odot$ for models
R15, B25, and W40, respectively.  In all cases, the
shock breakout luminosity exceeds the presupernova luminosity, with
characteristic values $\sim 10^{39}$, $10^{41}$, and $10^{42}$\,erg\,s$^{-1}$
for RSG, BSG, and WR, respectively.  The timescales range from days, hours, and
seconds, and the emission is dominated by optical, UV, and X-rays,
for RSGs, BSGs, and WRs, respectively. These estimates ignore the
effects of an intervening wind (e.g.,
\citealt{chevalier_2011,katz_2012,haynie_2020}), hence they are likely very
rough, particularly for WRs.  The EOS introduces a variation of a factor of a few in luminosity and
timescale, while the uncertainty in breakout velocity (Table~\ref{t:resolution}) 
adds another factor of $2$ uncertainty. Overall, resuls are qualitatively the same as those from
F18, as expected from Table~\ref{t:f18_comparison}

Plateau emission is estimated using the formulae of \citet{kleiser_2014}
\begin{eqnarray}
L_{\rm pl} & = & 1.2\times 10^{42} E_{\rm ej,51}^{5/6}M_{\rm ej,10}^{-1/2} R_{\rm cc,500}^{2/3}
             \kappa_{\rm 0.4}^{-1/3}T_{\rm rec,6000}^{4/3}\,\textrm{erg\,s}^{-1}\nonumber\\ 
&&\\
t_{\rm pl} & = & 120 E_{\rm ej,51}^{-1/6}M_{\rm ej,10}^{1/2} R_{\rm cc,500}^{1/6}                                           
             \kappa_{\rm 0.4}^{1/6}T_{\rm rec,6000}^{-2/3}\,\textrm{d},
\end{eqnarray}
where $L_{\rm pl}$ is the plateau luminosity and $t_{\rm pl}$ the plateau
duration, $E_{\rm ej,51} = E_{\rm ej}/(10^{51}$\,erg$)$, {$R_{\rm cc,500} =
R_{\rm cc}/(500R_\odot)$}, $\kappa_{\rm 0.4}$ is the opacity in units of
$0.4$\,cm$^2$\,g$^{-1}$, and $T_{\rm rec,6000}$ is the recombination
temperature at the ejecta photosphere in units of $6000$\,K (see also \citealt{kasen_2009}).  
We evaluate these quantities assuming a
recomination temperature of $10^{4}$\,K for RSGs and WRs, and $6000$\,K for
BSGs given the surface composition, as well as $\kappa_{\rm 0.4} = 1$ (F18).  The
characteristic expansion velocity of the ejecta is $v_{\rm exp} = \sqrt{2E_{\rm
ej}/M_{\rm ej}}$.  Table~\ref{t:observations} shows that these
properties are again consistent with those estimated in F18, with plateau emission lasting
for years, months, and days for RSGs, BSGs, and WRs, respectively. Luminosities
are much fainter than normal supernovae, with characteristic values $10^{38}$,
$10^{39}$, and $10^{37}$\,erg\,s$^{-1}$ for RSGs, BSGs, and WRs, respectively.
In the latter case, plateau emission is fainter than the presupernova luminosity.

In the context of modern optical transient surveys (e.g., \citealt{graham_2019}), 
the most promising signatures of non-rotating failed supernovae remain shock breakout in 
RSGs and plateau emission in BSGs, both of which have durations of the order of days 
and bolometric luminosities $10^{39}$\,erg\,s$^{-1}$. Shock breakouts from BSGs are bright 
candidates for surveys with hour-long cadences and UV capabilities.

The emergence of fast blue optical transients (e.g., \citealt{drout_2014}) 
as an observational class
has driven interest in failed supernovae from BSGs as possible progenitors  
(e.g., \citealt{kashiyama_2015}). Although the plateau luminosities by themselves
would place these transients at the faint end of the class \citep{suzuki_2020},
more power can be extracted from the compact object by extended fallback accretion 
from the ejected shell (F18), which would result in a brighter lightcurve
(e.g., \citealt{dexter_2013,moriya_2019}). Such a fallback scenario has
been proposed as a possible driver of activity from AT 2018cow \citep{margutti_2019}.
Additional power and/or diversity of properties can arise if
ejected shells intract with a dense circumstellar medium (e.g., \citealt{tsuna_2020}).

A dark collapse is still a possibility for some progenitors. Very small amounts of
bound ejecta are predicted for our YSGs and the most massive BSG
in our sample. Transients from these events might be faint enough that they are missed
by surveys, and collapse is only found in searches for disappearing progenitors 
(e.g., \citealt{reynolds_2015}).

\section{Summary and Discussion}
\label{s:summary}

We have carried out global simulations of non-rotating failed supernovae
in spherical symmetry, modeling the evolution of the inner supernova core 
($r\lesssim 10,000$\,km) with the general-relativistic, neutrino radiation-hydrodynamics
code {\tt GR1D}. The resulting gravitational mass loss is then used in
a Newtonian hydrodynamic simulation that follows the response of 
the outer layers of the star to the change in gravity. Relative to previous work by F18, 
we can now connect the EOS of dense matter to final ejecta masses and energies from 
failed supernovae and the associated electromagnetic emission. We also employ
much higher spatial resolution in the outer stellar layers than in previous work, 
thus addressing uncertainties in ejecta masses, energies, and velocities.
Our main results are the following:
\newline

\noindent
1. -- The ejecta masses and energies can vary by a factor of several depending
on the stiffness of the EOS. The dominant effect that determines this dependency
is the time to BH formation, which is longer for a stiffer EOS, resulting
in more gravitational mass lost to neutrinos (Fig.~\ref{f:dM_t}) and therefore a 
more energetic shock (Fig.~\ref{f:all_m4nr_E}). 

Previous work by F18, which used a parametric approach to the inner core evolution,
is consistent with our stiff EOS (DD2) results (Table~\ref{t:f18_comparison}) given 
the maximum NS mass ($2.5$\,$M_\odot$) used in their parametric scheme. 
Our results are also consistent with the trends with maximum 
NS mass found in the work of \citet{Lovegrove_Woosley_2013}, which also used
a parametric approach for the inner core evolution.
\newline

\noindent
2. -- When using a soft EOS (SFHo), our RSG and YSG progenitors fail to eject
unbound mass (Table~\ref{t:results}). Accounting for the energy released by 
hydrogen recombination (not included in the {\tt FLASH} EOS)
could provide sufficient thermal energy to unbind the ejeta in most of these bound 
cases.
\newline

\noindent
3. -- Predictions for shock breakout and plateau emission remain largely unchanged
relative to F18, with variations of a factor of a few in luminosity, duration,
velocity, and effective temperature (Table~\ref{t:observations}). The 
most promising candidates for optical transient surveys remain the shock
breakout from an RSG and plateau emission from a BSG, both of which have
durations on the order of days and luminosities on the order of $10^{39}$\,erg\,s$^{-1}$.
Detecting shock breakout from BSGs will require UV photometry on an hour-long cadence.
\newline

\noindent
4. -- Using a linear ramp with time for the gravitational mass lost to 
neutrinos (Fig.~\ref{f:W40S_m4nr_vs_m5_dM}) yields ejecta masses and energies within 
$\sim 10\%$ of those obtained using the detailed history of $\Delta M_{\rm G}(t)$
(Fig.~\ref{f:W40S_m4nr_vs_m5_EMv}). Variations in the neutrino luminosity on timescales 
smaller than the {BH} formation time have a very limited impact 
on mass ejection properties.
\newline

\noindent
5. -- Resolving the surface pressure scale height of the WR progenitors with a 
few cells results in an increase of $\sim 10\%$ in ejecta masses and energies 
relative to the unresolved case (Table~\ref{t:resolution}). Further increases
in resolution have the most impact when mass ejection is weak, as inferred
from our RSG and BSG models. Analytic estimates for
the shock breakout velocity agree to within a factor of $\sim 2$ with our numerically
measured values (Fig.~\ref{f:vbo_radius}).
\newline

\noindent
6. -- With our fully resolved stellar surfaces, we searched for the precursor
shocks predicted by \citet{coughlin_2018a}, finding clear evidence for it
in the BSG progenitor (Fig.~\ref{f:surface_shock}). A strong surface shock
is visible in the RSG, but inconsistencies in the thermodynamics ($T > 10^4$\,K
in the EOS used in {\tt FLASH}) preclude a more definitive statement. The WR progenitor
does not show a clear surface shock while displaying a noticeable positive
velocity for the entire photospheric region as the main shock approaches
the stellar surface. Expansion of the stellar surface prior to the arrival
of the main shock is only significant in our RSG models.
\newline

The main uncertainty for mass ejection and electromagnetic signal
prediction is the angular momentum distribution in the progenitor star.
If the infalling stellar material can circularize into an accretion disk,
outflows from this disk can eject additional matter or even reverse
the infall of stellar layers still in the process of collapsing 
(e.g., \citealt{woosley_2012,Quataert_Kasen_2012,Perna_et_al_2014,
kashiyama_2015,kashiyama_2018,murguia-berthier_2020,zenati_2020}).
General considerations about the angular momentum in presupernova envelopes 
suggest that disk formation is ubiquitous (F18), and even in the absence
of significant rotation, convective eddies in RSG envelopes can give rise 
to transient disk activity once they collapse \citep{quataert_2019}.

An additional uncertainty for electromagnetic signal prediction 
is the amount of mass present in the circumstellar medium of the progenitor.
In addition to the strong line-driven winds in WR stars, enhanced mass
loss prior to core-collapse is regularly inferred from supernova observations
(e.g., \citealt{bruch_2020}). Theoretically, enhanced mass loss (above steady 
line-driven winds) is expected from binary interactions (e.g. \citealt{smith_2014}) 
or from additional energy injection such was wave energy 
dissipation \citep{quataert_2012,quataert_2016,fuller_2017,fuller_2018}. 
This enhanced mass loss can significantly modify the shock breakout signal 
(e.g., \citealt{chevalier_2011,katz_2012,haynie_2020}).

Besides the outer structure of the progenitor, more precise estimates for the
strength of the main shock driven by gravitational mass loss can be obtained 
by using a general-relativistic treatment for the entire star, better neutrino transport, 
and more realistic progenitor models. 
Multi-dimensional simulations of BH-forming
supernovae can also predict additional features such as SASI-modulated neutrino
and/or gravitational wave activity (e.g., \citealt{pan_2020}, in particular
their non-rotating progenitor), which would complement the electromagnetic
signal in diagnosing the physics of stellar-mass BH formation, should
such an event occur in our Galaxy.


\acknowledgments

We thank Coleman Dean, Steven Fahlman, Craig Heinke, Sharon Morsink, Mathieu Renzo, 
and Greg Sivakoff for helpful discussions and/or comments on the manuscript.
We also thank Matthias Hempel for information about the DD2 EOS, and Alex Heger for information 
about the s20 and s40 progenitors.
{The anonymous referee provided constructive comments that improved the manuscript.}
This research was supported by the National Sciences and Engineering Research Council
(NSERC) of Canada through Discovery Grant RGPIN-2017-04286,
and by the Faculty of Science at the University of Alberta.
The software used in this work was in part developed by the U.S. Department
of Energy (DOE)
NNSA-ASC OASCR Flash Center at the University of Chicago.
Data visualization was done in part using {\tt VisIt} \citep{VisIt}, which is supported
by DOE with funding from the Advanced Simulation and Computing Program
and the Scientific Discovery through Advanced Computing Program.
This research was enabled in part by computing and storage support
provided by WestGrid (www.westgrid.ca), the Shared Hierarchical
Academic Research Computing Network (SHARCNET, www.sharcnet.ca),
Calcul Qu\'ebec (www.calculquebec.ca), and Compute Canada (www.computecanada.ca).
Computations were performed at the \emph{graham}, \emph{cedar}, and \emph{b\'eluga} clusters.

\software{{{\tt FLASH}} version 3 \citep{fryxell00,dubey2009},
          {{\tt GR1D}} version 1 \citep{oconnor_2010},
          {\tt Matplotlib} \citep{hunter2007},
          {{\tt MESA}} version 6794 \citep{paxton2011,paxton2013,paxton2015,paxton2018},
	  {\tt MESA SDK} \citep{mesa_sdk},
          {{\tt NumPy}} \citep{harris2020array},
          {{\tt VisIt}} \citep{VisIt}
          }


\bibliographystyle{aasjournal}
\bibliography{ms}

\label{lastpage}

\end{document}